\documentclass[showpacs,letterpaper,twocolumn,english,pra,aps]{revtex4}
\usepackage[T1]{fontenc}
\usepackage[latin1]{inputenc}
\usepackage{graphicx}
\usepackage{amssymb}

\makeatletter

\providecommand{\tabularnewline}{\\}

\large  

\input epsf

\usepackage{babel}
\makeatother
\begin{document}

\title{The Zel'dovich effect and evolution of atomic Rydberg spectra along the
Periodic Table}

\author{Eugene B. Kolomeisky and Michael Timmins}

\affiliation{Department of Physics, University of Virginia, 382 McCormick Rd.,
P. O. Box 400714, Charlottesville, VA 22904-4714}

\begin{abstract}
In 1959 Ya. B. Zel'dovich predicted that the bound-state spectrum
of the non-relativistic Coulomb problem distorted at small distances by
a short-range potential undergoes a peculiar reconstruction whenever
this potential alone supports a low-energy scattering resonance.
However documented experimental evidence of this effect has been
lacking.  Previous theoretical studies of this phenomenon were confined
to the regime where the range of the short-ranged potential is
much smaller than Bohr's radius of the Coulomb field.  We go beyond
this limitation by restricting ourselves to highly-excited $s$ 
states.  This allows us to demonstrate that along the Periodic Table of
elements the Zel'dovich effect manifests itself as systematic periodic 
variation of the Rydberg spectra with a period proportional to the
cubic root of the atomic number.  This dependence, which is supported by
analysis of experimental and numerical data, has its origin in the binding
properties of the ionic core of the atom. 

\end{abstract}


\maketitle

\section{Introduction}

In a variety of applications in physics it is important to understand
how is the normal Hydrogen spectrum modified if at small distances
the Coulomb law is replaced by a central short-ranged potential. An
important aspect of this problem is the existence of two length scales
- the Bohr radius of the Coulomb field $a_{B}$ and the range of action
of short-range forces $r_{0}$.

For example, in hadronic atoms formed by charged particles and antiparticles
the large distance Coulomb attraction gives its way at short distances
to nuclear forces whose range $r_{0}$ is significantly smaller than
$a_{B}$ \cite{LL1}. 

In condensed matter physics a similar problem is that of the energy
spectrum of the Wannier-Mott exciton \cite{AM}. When in a semiconductor
an electron is excited into the conduction band, a bound state with
a hole left in the valence band can form. Due to the large dielectric
constant of the medium the electron and the hole in the exciton are
spatially well-separated. Therefore the electron-hole interaction
is a Coulomb attraction modified at short distances. In this context
$a_{B}$ can exceed many times $r_{0}$ which is of the order of the
Hydrogen Bohr radius. 

Zel'dovich was apparently the first to recognize that in the limit
$r_{0}\ll a_{B}$ the spectrum of the distorted Coulomb problem is
peculiar \cite{Zel'dovich}. Since the centrifugal barrier decreases
the probability of particle penetration in the region of small
distances $r$, the effect of the short-range potential is strongest
for the states of zero angular momentum. In this case the radial motion
of a particle of mass $m$ and energy $E$ in a central potential
$U(r)$ is described by the one-dimensional Schr\"{o}dinger equation
\cite{LL2}\begin{equation}
{\frac{d^{2}{\chi}}{dr^{2}}}+{\frac{2m}{\hbar^{2}}}\left(E-U(r)\right)\chi=0\label{SE}\end{equation}
 where $\chi(r)/r$ is the radial wave function. Zel'dovich chose
$U(r)=-\hbar^{2}/ma_{B}r$ for $r\geq r_{0}$ and $U(r)=U_{s}(r)$
otherwise and demonstrated that as long as the short-range potential
$U_{s}(r)$ is not resonant, its effect is weak. If, on the other
hand, $U_{s}(r)$ has a low-energy scattering resonance, a drastic
reconstruction of the spectrum takes place. Using the example of the
square well of depth $U_{0}$ he stated that as the dimensionless
coupling constant $w\simeq mr_{0}^{2}U_{0}/\hbar^{2}$ increases,
the spectrum of the problem $E_{n}(w)$ evolves in a fashion resembling
a sharp decreasing staircase. The steps are located at critical values
of $w$ at which bound states occur in $U_{s}(r)$ \textit{only}.
As $w$ goes through the first threshold, the Coulomb levels $E_{n}$
($n\geq2$) quickly fall to $E_{n-1}$ while the ground state $E_{1}$
rapidly drops downward. The relative width of the region where the
spectrum reconstruction takes place, $\Delta w/w\simeq r_{0}/a_{B}\ll1$,
is narrow, and qualitatively the same pattern repeats itself upon
passing through every subsequent resonance.

A similar spectral behavior has been found by Popov \cite{Popov1}
in his analysis of the Dirac equation for an electron in a field of
the bare nucleus of charge $Ze$ with $Z>137$. 

The Zel'dovich effect has been re-discovered in the spectra of hadronic
atoms, and its generality has been demonstrated for any interaction
with two widely different spatial scales \cite{Shapiro}.

Various aspects of the spectrum reconstruction have been investigated
by Popov and collaborators \cite{Popov2}. Their study was motivated
by then existing experimental evidence of the large $1s$-level shift
in the proton-antipropton atom which was naturally linked to the Zel'dovich
effect. Later it became clear that the experimental level shifts are
small and the interest in the phenomenon declined.

As far as we know, at this time there is no documented experimental
evidence of the Zel'dovich effect. This is not surprising because
the spectrum reconstruction takes place in a narrow range of parameters
in the vicinity of low-energy resonances. However a given experimental
system is unlikely to be near resonance. A systematic search for the
Zel'dovich effect would consist in looking for spectral changes in
response to tuning of the central part of the potential which is often
impossible - the strength of the nuclear force cannot be changed in
the laboratory.

Recently Karnakov and Popov \cite{KarnakovPopov} pointed out that
the Zel'dovich spectrum reconstruction takes place for a Hydrogen
atom as a function of the external magnetic field thus providing an
example of a system where a systematic search for the effect might be
possible. Although the phenomenon is observable in numerical studies,
direct experimental evidence is lacking and may only come from astrophysical
observations as the pertinent magnetic fields are comparable to those
on the surface of a neutron star.

The goal of this paper is to demonstrate that evolution of the Rydberg
spectra of ordinary atoms along the Periodic Table provides direct
evidence of the Zel'dovich effect. Since the condition $r_{0}\ll a_{B}$
does not hold in atomic systems, the way the phenomenon manifests
itself is less dramatic - we will show that it can be seen as a systematic
periodic spectral modulation as a function of the cubic root of atomic number
$Z$.

It is known that for a highly excited $s$ electron of a Rydberg atom
the effect of polarization of the ionic core is negligible compared
to that of the wave function penetration in the central region of
the atom \cite{Friedrich}. Therefore the electron dynamics can be
adequately described by Eq.(\ref{SE}) where the effective central
field $U(r)$ at large distances is that of a positively charged ion
of charge $e$. On the other hand, starting from distances of
the order of the size of the ionic core $r_{0}\simeq a_{B}$ the field
felt by the electron begins to deviate from the $-e^{2}/r$ form on-average
decreasing, as $r\rightarrow0$, to $-Ze^{2}/r$. By increasing $Z$
along the Periodic Table Nature systematically deepens the inner
part of the potential leaving the outer $-e^{2}/r$ tail intact. Thus
by analyzing the Rydberg spectra as a function of atomic number $Z$
it may be possible to correlate them with the binding properties of
the ionic core which will constitute evidence of the Zel'dovich
effect.

In atomic physics the motion of an electron in the field of a residual
atomic ion has been studied in the past. Approximating the potential
of the ionic core by that of the Thomas-Fermi or Thomas-Fermi-Dirac
theories Latter \cite{Latter} computed numerically the single-electron
term values from $1s$ to $7d$ for all atoms. His $ns$ spectra as
a function of atomic number $Z$ for largest $n$ studied clearly
show modulations on a decreasing energy curve. It is well-known that
the large $n$ atomic spectra are described by the Rydberg formula
\cite{Friedrich}\begin{equation}
E_{n}=-{\frac{\hbar^{2}}{2ma_{B}^{2}}}{\frac{1}{(n-\mu)^{2}}}\label{Rydberg}\end{equation}
 where $\mu$ is the quantum defect which in the limit $n\rightarrow\infty$
does not depend on $n$. Latter's results imply that the dependence
of the quantum defect $\mu$ on $Z$ has modulations superimposed
on an increasing curve.

The $\mu(Z)$ dependence has been numerically computed by Manson \cite{Manson}
and by Fano, Theodosiou and Dehmer \cite{Fano} who used the Hartree-Slater
model \cite{HS} to approximate the potential of the ionic core of the
atom. Although the periodic variations of $\mu(Z)$ are strongly obscured
by the shell effects (included in the Hartree-Slater model), Fano,
Theodosiou and Dehmer argued that they are there and that there is
a correlation between the location of radial nodes of the function
$\chi$ from Eq.(\ref{SE}) near $r_{0}$ and the slope of the $\mu(Z)$
dependence. In view of the oscillation theorem \cite{LL2} the nodal
structure of the function $\chi$ is intimately related to the binding
properties which suggests that systematic periodic variations of Rydberg
spectra as function of $Z$ might be related to the Zel'dovich effect.

In order to demonstrate that this connection is correct below we compute
the upper part of the spectrum of the modified Coulomb problem not
assuming that $r_{0}\ll a_{B}$. We show that the staircase reconstruction
taking place for $r_{0}\ll a_{B}$ and the spectral modulations for
$r_{0}\simeq a_{B}$ are different limiting cases of the same phenomenon
- sensitivity to the binding properties of the inner part of the potential
which we continue to call the Zel'dovich effect. We also compare our
results for $\mu(Z)$ with available experimental and numerical data
to show that the phenomenon is observable.

The organization of this paper is as follows.  In Section II we
provide a short derivation of the Rydberg formula (\ref{Rydberg}) and
arrive at the expression for the quantum defect in terms of
the dimensionless range of the inner potential and its scattering length.
This general result is further analyzed in the $r_{0} \ll a_{B}$
limit and the main features of the Zel'dovich spectral reconstruction
are recovered (Section IIA).  In Section IIB we establish a
relationship between the Zel'dovich effect and Levinson's theorem of 
quantum mechanics.  This is followed (Section IIC) by the analysis of
the opposite $r_{0} \gg a_{B}$ limit where we demonstrate that the
Zel'dovich effect manifests itself in the form of a spectral
modulation whose origin still lies in the binding properties of the inner
potential $U_{s}(r)$.  These general findings are illustrated in
Section IID where we use the exactly-solvable example of the
rectangular well as a model for the inner potential.   In Section IIE
we observe that only a treatment more accurate than semiclassical can
capture the Zel'dovich effect. 

Section III focuses on the computation of the systematic quantum defect of
the Rydberg electron as a function of atomic number $Z$.  First 
(Section IIIA), for the inner potential having an attractive Coulombic
singularity at the origin, we derive a semiclassical expression for the
quantum defect and show that it is equal to the number of de Broglie's
half-waves fitting inside the inner potential minus a contribution
proportional to $(r_{0}/a_{B})^{1/2}$.  Going beyond the semiclassical
approximation we also demonstrate that the Zel'dovich modulation of the
quantum defect is a periodic function of the number of de Broglie's
half-waves fitting inside the ionic core of the atom.  This is followed
by an explicit calculation based on Latter's model of the ionic core 
\cite{Latter}.  First, the semiclassical quantum defect is calculated
as a function of $Z^{1/3}$ (Section IIIB).  Then (in Section IIIC) a
full computation capturing the Zel'dovich effect is performed.  An
important ingredient here is an approximate calculation of the
scattering length of the ionic core of the atom.  Both the scattering
length and the related Zel'dovich modulation of the quantum defect
turned out to be nearly periodic functions of $Z^{1/3}$.   

In Section IV the results of our systematic calculation are compared
with experimental and numerical data.  First, we observe that the bulk
of the quantum defect values is well-captured semiclassically.  Then 
(Section IVA) we demonstrate that the gross features of the deviation
away from semiclassics are due to the effects of the shell structure. 
This is done by  establishing and demonstrating a correlation between
the variation of the radius of the ionic core of the atom and 
corresponding variation of the quantum defect. Finally, in Section IVB
a Fourier analysis of the quantum defect variation with $Z^{1/3}$
is conducted which singles out the Zel'dovich effect.  As a by-product
we also find a $Z^{1/3}$ periodic contribution coming from the shell effects.  

We conclude (Section V) by outlining our main result and directions of
future work. 

\section{Distorted Coulomb problem and quantum defect}

We will be interested in the low energy bound states with the classical
turning point being far away from the boundary of the central region,
i. e. $\hbar^{2}/ma_{B}|E|\gg r_{0}$. Then the quickest way to derive
the spectrum is via semiclassical arguments derived from those given
by Migdal \cite{Migdal}: 

For $r_{0}<r<\hbar^{2}/ma_{B}|E|$ the semiclassical solution to Eq.(\ref{SE})
can be written in two equivalent forms: 
\begin{eqnarray}
\label{scfunction}
\chi_{sc} & \propto & {\frac{1}{\sqrt{p}}}\sin\left({\frac{1}{\hbar}}\int\limits _{r}^{\hbar^{2}/ma_{B}|E|}pdr+{\frac{\pi}{4}}\right)\nonumber \\
 & \propto & {\frac{1}{\sqrt{p}}}\sin\left({\frac{1}{\hbar}}\int\limits _{r_{0}}^{r}pdr+\alpha\right)\end{eqnarray}
 where $p=(-2m|E|+2\hbar^{2}/ra_{B})^{1/2}$ is the momentum. The
first representation in Eq.(\ref{scfunction}) is the standard result
with the phase of $\pi/4$ improving on the deficiency of the semiclassical
approximation near the classical turning point, while the yet undetermined
phase $\alpha$ in the second representation in Eq.(\ref{scfunction})
both corrects for the failure of the semiclassical approximation in
a Coulomb field at distances $r \lesssim a_{B}$ and accounts for the
short-range potential $U_{s}(r)$.

For $r_{0}<r\ll\hbar^{2}/ma_{B}|E|$ the Schr\"{o}dinger equation
(\ref{SE}) simplifies to \begin{equation}
{\frac{d^{2}{\chi}}{dr^{2}}}+{\frac{2}{ra_{B}}}\chi=0\label{E=0SE}\end{equation}
 and can be exactly solved: \begin{equation}
\chi\propto r^{1/2}\left(J_{1}(\sqrt{8r/a_{B}})-Y_{1}(\sqrt{8r/a_{B}})\tan\delta\right)\label{E=3D0function}\end{equation}
 where $J_{\nu}(x)$ and $Y_{\nu}(x)$ are the order $\nu$ Bessel
functions of the first and second kind respectively \cite{Lebedev}.
The solution (\ref{E=3D0function}) is a linear combination of the
regular $J_{1}(0)=0$ and irregular $Y_{1}(0)=\infty$ Coulomb functions
of zero energy, and for the purely Coulomb problem, 
$U_{s}(r)=-\hbar^{2}/ma_{B}r$, one has to recover $\tan\delta=0$.

For $a_{B}\ll r\ll\hbar^{2}/ma_{B}|E|$ the semiclassical approximation
is accurate, and the second representation of Eq.(\ref{scfunction})
yields $\chi\propto r^{1/4}\sin(\sqrt{8r/a_{B}}-\sqrt{8r_{0}/a_{B}}+\alpha)$.
On the other hand, the $r\gg a_{B}$ limit of (\ref{E=3D0function})
is $\chi\propto r^{1/4}\sin(\sqrt{8r/a_{B}}-\pi/4+\delta)$ which determines
$\alpha$ in (\ref{scfunction}) to be $\sqrt{8r_{0}/a_{B}}-\pi/4+\delta$.
It also implies that $\delta$ in (\ref{E=3D0function}) is the zero-energy
phase shift due to the small-distance deviation of the potential from
the Coulomb form.

The energy spectrum can be found from the requirement that the semiclassical
expressions (\ref{scfunction}) coincide. Combined with $\alpha=\sqrt{8r_{0}/a_{B}}-\pi/4+\delta$
this gives the quantization rule \begin{equation}
{\frac{1}{\hbar}}\int\limits _{r_{0}}^{\hbar^{2}/ma_{B}|E|}pdr=\pi n-\delta-x_{0}\label{quantization}\end{equation}
 where dimensionless parameter \begin{equation}
x_{0}=\sqrt{\frac{8r_{0}}{a_{B}}}\label{x0}\end{equation}
 measures the range of the short-range forces. Calculating the integral
we arrive at Eq.(\ref{Rydberg}) with $\delta=\pi\mu$ which is the
statement of Seaton's theorem \cite{Friedrich} relating the quantum
defect to the zero-energy phase shift.

The range of applicability of Eq.(\ref{Rydberg}), $n-\mu\gg(r_{0}/a_{B})^{1/2}\simeq x_{0}$,
follows from the condition $|E|\ll\hbar^{2}/ma_{B}r_{0}$ which also
implies that in order to calculate the quantum defect entering the
spectrum (\ref{Rydberg}), we only need to match (\ref{E=3D0function})
with its zero energy counterpart at $r<r_{0}$.

We proceed by computing $h=[d\ln\chi/d\ln r]_{r\rightarrow r_{0}+0}$,
the logarithmic derivative of the function (\ref{E=3D0function})
evaluated at the boundary of the inner region: \begin{equation}
h={\frac{x_{0}}{2}}{\frac{J_{0}(x_{0})-Y_{0}(x_{0})\tan\pi\mu}{J_{1}(x_{0})-Y_{1}(x_{0})\tan\pi\mu}}\label{logderivative}\end{equation}
 where we used $\delta=\pi\mu$. The quantum defect $\mu$ is determined
by equating (\ref{logderivative}) to $h_{s}=[d\ln\chi/d\ln r]_{r\rightarrow r_{0}-0}$
which can be found by solving the $E=0$ Schr\"{o}dinger equation
(\ref{SE}) for $r<r_{0}$ with $U(r)=U_{s}(r)$: \begin{equation}
{\frac{d^{2}{\chi}}{dr^{2}}}-{\frac{2m}{\hbar^{2}}}U_{s}(r)\chi=0\label{zeroSE}\end{equation}
 The parameter $h_{s}$ can be equivalently expressed in terms of
the scattering length corresponding to the inner potential \textit{only}.
Indeed for motion in a short-range potential the scattering length
$a_{s}$ is defined from the asymptotic $r\rightarrow\infty$ behavior
$\chi(r)\propto 1-r/a_{s}$ of the solution to (\ref{zeroSE}). For a potential
well identically vanishing for $r>r_{0}$, this is also the exact
behavior outside the well with the implication that \cite{Zel'dovich,Popov2}\begin{equation}
h_{s}^{-1}=\left({\frac{d\ln\chi(r\rightarrow r_{0}-0)}{d\ln r}}\right)^{-1}=1-{\frac{a_{s}}{r_{0}}}\label{logderscatteringlength}\end{equation}
 Then substituting $h=h_{s}$ in Eq.(\ref{logderivative}) and using
(\ref{logderscatteringlength}) we arrive at the formula for the quantum
defect \begin{equation}
\tan\pi\mu={\frac{2x_{0}^{-1}J_{1}(x_{0})+(a_{s}/r_{0}-1)J_{0}(x_{0})}{2x_{0}^{-1}Y_{1}(x_{0})+(a_{s}/r_{0}-1)Y_{0}(x_{0})}}\label{qdgeneral}\end{equation}
 If the short-distance potential is selected in the form $U_{s}(r)=-\hbar^{2}/ma_{B}r$,
i. e. we have the ordinary Coulomb problem in the whole space, the
quantum defect $\mu$ entering the Rydberg formula (\ref{Rydberg})
must vanish identically. It is straightforward to verify that this
is indeed the case: the $E=0$ inner $r<r_{0}$ solution to (\ref{SE}),
$\chi\propto r^{1/2}J_{1}(\sqrt{8r/a_{B}})$, leads to the expression
for the scattering length nullifying the numerator of (\ref{qdgeneral}).
This argument defines the zero of the quantum defect and implies that
$\mu$ is necessarily positive if for all $r\leq r_{0}$ the inner
potential $U_{s}(r)$ is more attractive than the Coulomb potential
$-\hbar^{2}/ma_{B}r$; otherwise the quantum defect is negative. For
example, for $U_{s}(r)=0$ the quantum defect $\mu(0,x_{0})$ is a
negative monotonically decreasing function of $x_{0}$ such as $\mu=-x_{0}^{4}/32$
for $x_{0}\ll1$, and $\mu=3/4-x_{0}/\pi$ in the opposite $x_{0}\gg1$
limit. The $\mu(0,x_{0})$ dependence as well as its $x_{0}\gg1$
limit are shown in Fig.~1. 
\begin{figure}

\includegraphics[
  width=1.0\columnwidth,
  keepaspectratio]{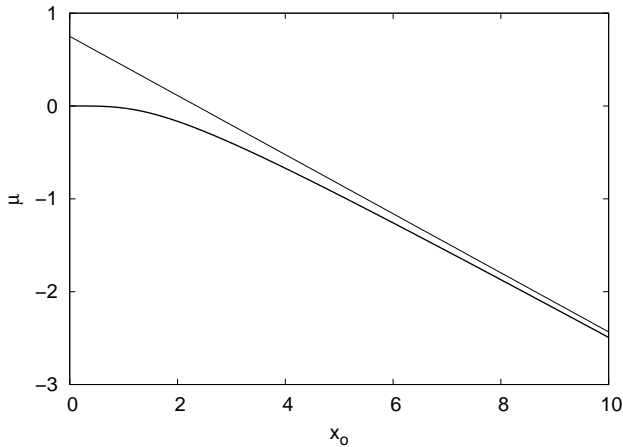}

\caption{Quantum defect for $U_{s}(r)=0$ as a function of the range parameter
$x_{0}$, Eq.(\ref{x0}), and its $x_{0}\gg1$ limit, $\mu(0,x_{0})=3/4-x_{0}/\pi$
(shown in gray scale).}

\end{figure}

\subsection{Zel'dovich effect in the $r_{0}\ll a_{B}$ limit}

For $x_{0}\ll1$ Eq.(\ref{qdgeneral}) simplifies to a form accumulating
the physics of the Zel'dovich effect: \begin{equation}
{\frac{a_{B}}{2\pi a_{s}}}\equiv{\frac{4}{\pi x_{0}^{2}(1-h_{s}^{-1})}}=-\cot\pi\mu-{\frac{2}{\pi}}\ln{\frac{2}{\gamma x_{0}}}\label{qdzel'dovichlimit}\end{equation}
 where $\ln\gamma=0.5772$ is Euler's constant. Terms of higher order
in $x_{0}$ which for $U_{s}(r)=0$ lead to small negative values of
the quantum defect are neglected in (\ref{qdzel'dovichlimit}).

We verified that Eq.(\ref{qdzel'dovichlimit}) matches the upper portion
of the $ns$ spectrum which for $r_{0}\ll a_{B}$ is known in closed
form for any $n$ \cite{Popov2}. We also note that with some effort
Eq.(\ref{qdzel'dovichlimit}) can be deduced from the expression for
the phase shift of the proton-proton scattering given by Landau and
Smorodinskii \cite{LS}: in their formula we have to (i) reverse the
sign of the Bohr radius, (ii) take the limit of zero energy, and (iii)
employ Seaton's theorem \cite{Friedrich} $\delta=\pi\mu$. 
\begin{figure}
\includegraphics[
 width=1.0\columnwidth,
 keepaspectratio]{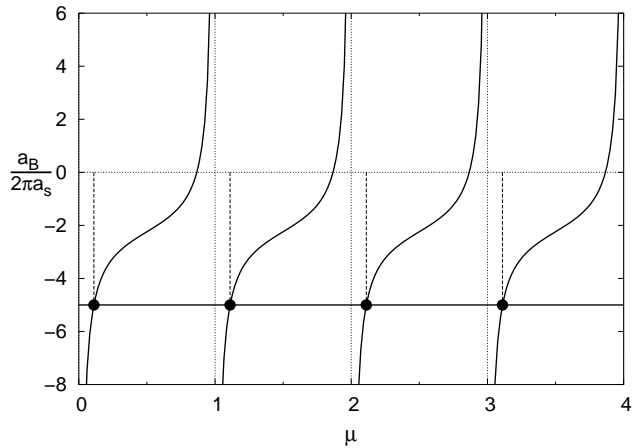}

\caption{Graphical solution of Eq.(\ref{qdzel'dovichlimit}); $x_{0}=1/30$
has been used to construct the graph. The quantum defect $\mu$ is
given by the intersections of the right-hand side of (\ref{qdzel'dovichlimit})
with the line of constant $a_{B}/2\pi a_{s}$.}

\end{figure}

Fig.~2 shows the dependence of $a_{B}/2\pi a_{s}$ on $\mu$ given
by Eq.(\ref{qdzel'dovichlimit}); its inverse $\mu(a_{B}/2\pi a_{s})$
is a multivalued function consisting of a series of increasing step-like
curves sandwiched between nearest non-negative integers. The slope
of $\mu(a_{B}/2\pi a_{s})$ is small everywhere except for the vicinity
of half-integer $\mu$.

Since typically the central potential $U_{s}(r)$ is not resonant,
$|h_{s}|$ in (\ref{qdzel'dovichlimit}) is not small. Then the magnitude
of the scattering length is of the order of the size of the inner
well, $|a_{s}|\simeq r_{0}$ and $a_{B}/2\pi|a_{s}|\simeq1/x_{0}^{2}$
is significantly larger than the last term in (\ref{qdzel'dovichlimit}).
This implies that the quantum defect is very close to an integer,
$\mu=-2a_{s}/a_{B}~(mod~1)$, with $|a_{s}|/a_{B}\simeq r_{0}/a_{B}\ll1$.
This conclusion is in quantitative agreement with the results of 
perturbation theory in $a_{s}/a_{B}$ when the deviation from the
Bohr Hydrogen formula is small \cite{Thirring,Zel'dovich}. It is
applicable to an attractive non-resonant well of arbitrary strength;
for weak $U_{s}(r)$ which cannot support a bound state we have $\mu=-2a_{s}/a_{B}>0$
\cite{note1} represented by the leftmost intersection in Fig.~2.
We also note that the spectrum is exactly Hydrogenic if the scattering
length is zero which can be viewed as an analog of the Ramsauer effect
\cite{LL3}: in the present context it refers to a resonant phenomenon
when the distortion of the Coulomb potential at small distances is
invisible to the low-energy bound (or incident) particle.

Exactly at half-integer $\mu$ the scattering length is negative with
the magnitude $|a_{s}|=(a_{B}/4)\ln^{-1}(2/\gamma x_{0})=2r_{0}x_{0}^{-2}\ln^{-1}(2/\gamma x_{0})$
significantly exceeding the size of the central region $r_{0}$. This
implies that the slope of the $\mu(a_{B}/2\pi a_{s})$ dependence
is largest when $U_{s}(r)$ itself is almost resonant so that it supports
a low-energy \textit{virtual} state. At the point of the steepest
slope we also have $[d\chi/dr]_{r=r_{0}}\propto h_{s}\simeq(x_{0}^{2}/2)\ln(2/\gamma x_{0})\ll1$.
Since this is practically zero, one can equivalently say that the
slope of the $\mu(a_{B}/2\pi a_{s})$ dependence is largest when the
antinode of the function $\chi$ in Eq.(\ref{SE}) occurs at the boundary
of the inner region $r_{0}$. This criterion resembles that given
by Fano, Theodosiou and Dehmer \cite{Fano} for the dependence of
the quantum defect $\mu$ on atomic number $Z$. We note however,
that for a Rydberg atom the size of the residual ion does not satisfy
the condition $r_{0}\ll a_{B}$; this issue is further addressed below.

If for all $r$ the central well is attractive, its effect can be
quantified by a single dimensionless coupling constant $w\simeq mr_{0}^{2}|U_{s}|/\hbar^{2}>0$
where $|U_{s}|$ has a meaning of the characteristic depth of the
well. Then the inverse scattering length $a_{s}^{-1}$ is known to
be a monotonically increasing function of $w$ \cite{Popov2} - an
$a_{s}$ dependence shown in Fig.~3 in gray scale is typical and
may help illustrate the argument given below.

The step-like features of the function $\mu(a_{B}/2\pi a_{s})$ are
amplified in the $\mu(w)$ dependence. Indeed, for $w\ll1$ the scattering
length $a_{s}$ is very small and negative. Then the line of constant
$a_{B}/2\pi a_{s}$ in Fig.~2 lies at very large negative values,
the quantum defect satisfies $\mu=-2a_{s}/a_{B}\ll1$, and the deviation
from the normal Hydrogen spectrum is small. As the well deepens, the
coupling constant $w$ increases, the scattering length becomes more
negative and the horizontal line of constant $a_{B}/2\pi a_{s}$ moves
upward. However as long as the well remains non-resonant, the quantum
defect $\mu$ will only grow very little. The strongest increase of
$\mu(w)$ in response to deepening of the well (and thus the largest deviation
from the Bohr Hydrogen formula) occurs when the scattering length
reaches a very large negative value $a_{s}=-2r_{0}x_{0}^{-2}\ln^{-1}(2/\gamma x_{0})$.
For $x_{0}\ll1$ this takes place very close to a threshold value
of the coupling constant $w$ when the first bound state is about
to appear in $U_{s}(r)$. The relative width of the reconstruction
region $\Delta w/w$ centered around $\mu=1/2$ thus can be estimated
from the scaling behavior of the scattering length near the threshold
$a_{s}\simeq r_{0}w/\Delta w$ and the condition $a_{B}\simeq|a_{s}|$.
This leads to the original result of Zel'dovich \cite{Zel'dovich}
$\Delta w/w\simeq r_{0}/a_{B}$.

As the coupling constant $w$ increases through the first binding
threshold, the inverse scattering length changes sign, and the line
of constant $a_{B}/2\pi a_{s}$ in Fig.~2 enters the region of positive
values. After passing through the reconstruction region, the positive
scattering length decreases in magnitude, for $a_{s}/a_{B}\ll1$ the
quantum defect is close to unity, $\mu=1-2a_{s}/a_{B}$, and the deviation
from the normal Hydrogen spectrum is again small. In the region $a_{s}\simeq r_{0}$
the scattering length does not vary strongly with the depth of the
well, and one can say that the slope of the $\mu(w)$ dependence will
be minimal when the node of the function $\chi$ in Eq.(\ref{SE})
is near the boundary of the central region $r_{0}$ which parallels
the criterion of Fano, Theodosiou and Dehmer \cite{Fano}. Upon further
increase of the coupling constant $w$, the scattering length gets
smaller and the line of constant $a_{B}/2\pi a_{s}$ in Fig. 1 enters
the region of very large positive values becoming infinite at $a_{s}=0$.

\begin{figure}
\includegraphics[
 width=1.0\columnwidth,
 keepaspectratio]{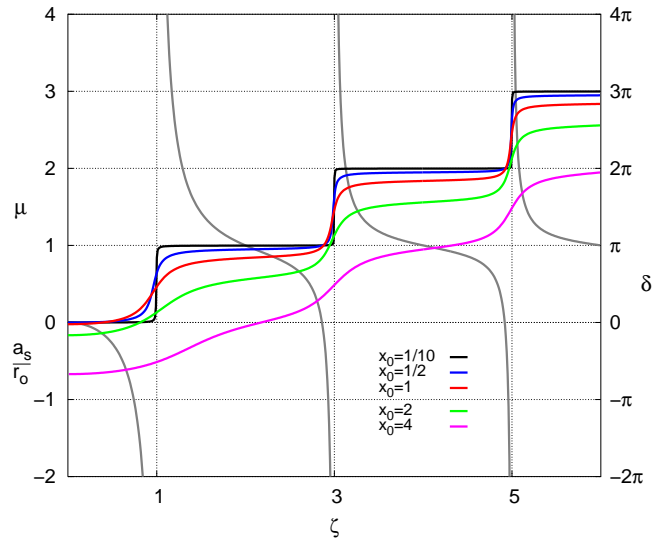}

\caption{Evolution of the Zel'dovich effect for the rectangular well of radius
$r_{0}$ and depth $U_{0}$ for a series of range parameters $x_{0}$,
Eq.(\ref{x0}), manifested in the dependences of the quantum defect
$\mu$ on $\zeta=(8mU_{0}r_{0}^{2}/\pi^{2}\hbar^{2})^{1/2}\simeq w^{1/2}$.
The correlation with the binding properties of the well is seen from
the plot of the reduced scattering length $a_{s}/r_{0}$ (gray scale).
Shown are also the values of the zero-energy phase shift $\delta=\pi\mu$
relating the Zel'dovich effect to Levinson's theorem.}

\end{figure}

To summarize, as $a_{s}(w)$ goes through one complete cycle decreasing
from zero, passing through the binding resonance, and then approaching
zero from above, the quantum defect $\mu(w)$ increases from zero
to unity in a staircase fashion: it is mostly zero or unity except
for the narrow region $\Delta w/w\simeq r_{0}/a_{B}\ll1$, $\mu\simeq1/2$
near the first binding threshold of $U_{s}(r)$. Combined with the
Rydberg formula (\ref{Rydberg}) this implies that the Coulomb levels
$E_{n}$ quickly fall to $E_{n-1}$ which constitutes the essence
of the Zel'dovich effect \cite{Zel'dovich}.

As the coupling constant $w$ continues to increase away from $a_{s}(w)=0$,
the next cycle, $1<\mu(w)<2$, begins and qualitatively same pattern
repeats itself. This remains true for every subsequent cycle with
$\mu(w)$ sandwiched between nearest integers. Overall the quantum
defect is an increasing function of $w$ having the form of a staircase
with practically integer plateaus and sharp steps located at half-integer
$\mu$. The steps correspond to the presence of the low-energy scattering
resonances in $U_{s}(r)$.

To illustrate this behavior we choose the inner potential in the form
of a rectangular well of depth $U_{0}$ whose scattering length is
given by $a_{s}/r_{0}=1-2\tan(\pi\zeta/2)/\pi\zeta$ with dimensionless
parameter $\zeta=(8mU_{0}r_{0}^{2}/\pi^{2}\hbar^{2})^{1/2}\simeq w^{1/2}$
quantifying the depth of the well. The scattering length diverges
at odd values of $\zeta$ which correspond to consecutive occurrences
of bound states in the well; the respective dependence of $a_{s}/r_{0}$
on $\zeta$ is shown in Fig.~3 in gray scale. We also plot the dependences
of the quantum defect $\mu$ on $\zeta$ found from the general expression
(\ref{qdgeneral}) for a series of representative $x_{0}$. The analysis
based on Eq.(\ref{qdzel'dovichlimit}) is illustrated by the $x_{0}=1/10$
and $x_{0}=1/2$ curves; the latter corresponds to the case of the
proton-antiproton atom \cite{Popov2}. These dependences have the form
of staircases with nearly integer plateaus; the steepness of the steps
where the quantum defect varies by unity and the flatness of the plateaus
increase as $x_{0}$ gets smaller. An inspection reveals that the
points of maximal slope of $\mu(\zeta)$ somewhat precede the scattering
resonances in accordance with the analysis given above. This is seen
most clearly for the $\zeta\simeq1$ step of the $x_{0}=1/2$ curve.
Fig.~2 of Zel'dovich's work \cite{Zel'dovich} has this feature as
well. From a practical standpoint the steps can be considered to coincide
with the binding resonances of the well.

The relative width of the reconstruction region $\Delta\zeta/\zeta$
can be estimated as $r_{0}/a_{B}\zeta^{2}$. Since the threshold values
$\zeta$ grow linearly with the number of bound states, then for fixed
$x_{0}$ the steepness of the steps increases with $\zeta$ as can
be seen in Fig.~3. This is merely the consequence of the sharpening
of the binding resonances. Similarly the flatness of the plateaus
improves as $\zeta$ increases, and the points of least slope of the
$\mu(\zeta)$ dependence asymptotically approach even values of $\zeta$.
This is where the node of the function $\chi$ in Eq.(\ref{SE}) coincides
with the boundary of the central region, $a_{s}=r_{0}$.

Finally we note that the quantum defect takes on exactly integer values
whenever the scattering length vanishes.

\subsection{Connection to Levinson's theorem}

There is a deep parallel between the Zel'dovich reconstruction of
the upper $E\rightarrow0$ part of the Coulomb spectrum in the $r_{0}\ll a_{B}$
limit and the low-energy scattering by a short-range potential well.
For a particle of energy $E=\hbar^{2}k^{2}/2m$ whose wave vector
${\textbf{k}}$ is small in magnitude, $kr_{0}\ll1$, scattered by
the short-range potential $U_{s}(r)$ vanishing for $r>r_{0}$ the
scattering length $a_{s}$ can be defined \cite{LL4} through the
$k\rightarrow0$ limit of the relationship \begin{equation}
1/ka_{s}=-\cot\delta_{s}(k)\label{phaseshift}\end{equation}
 where $\delta_{s}(k)$ is the phase shift. Employing Seaton's theorem
\cite{Friedrich} $\delta=\pi\mu$ relating the quantum defect to
the zero-energy phase shift it is straightforward to realize that
Eqs.(\ref{qdzel'dovichlimit}) and (\ref{phaseshift}) are direct
analogs. The Coulomb field is characterized by its own length scale,
the Bohr radius $a_{B}$. Its free particle counterpart entering Eq.(\ref{phaseshift})
is the de Broglie wavelength $2\pi/k$. The range of applicability
of Eq.(\ref{qdzel'dovichlimit}) $r_{0}\ll a_{B}$ parallels the low-energy
condition $kr_{0}\ll1$ necessary for Eq.(\ref{phaseshift}) to hold.
The analysis which led to the explanation of the Zel'dovich effect
can be repeated for Eq.(\ref{phaseshift}) with the conclusion that
the phase shift $\delta_{s}(k)$ as a function of the dimensionless depth
$w$ of the scattering well has the form of a sharp increasing staircase
whose plateaus practically coincide with $\delta_{s}(k)=0~(mod~\pi)$.
The steps where the phase shift changes by $\pi$ are very narrow,
$\Delta w/w\simeq kr_{0}\ll1$, and the points of steepest slope are
located at $\delta_{s}(k)\simeq\pi/2~(mod~\pi)$. In the limit $k\rightarrow0$
the staircase becomes perfect. This can be recognized as Levinson's
theorem \cite{LL4} relating the number of bound states in a well
with the zero-energy scattering phase shift. We conclude that for
$r_{0}\ll a_{B}$ the Zel'dovich effect expressed in terms of the
zero-energy phase shift $\delta$ is the Coulombic cousin of Levinson's
theorem \cite{Spruch}. A special case of this correspondence, the Ramsauer-like
recovery of the normal Hydrogen spectrum for $a_{s}=0$, was already
mentioned earlier. In the limit $x_{0}=\sqrt{8r_{0}/a_{B}}\rightarrow0$
Zel'dovich's staircase becomes perfect and identical to Levinson's
staircase. This can be understood as a result of taking the neutral
limit, $a_{B}\rightarrow\infty$, when the Coulomb part of the binding
potential $U(r)$ in (\ref{SE}) vanishes. From this viewpoint, Levinson's
theorem is a consequence of the Zel'dovich effect. To emphasize the
connection to Levinson's theorem, in Fig.~3 we additionally show
the zero-energy phase shift $\delta=\pi\mu$.

Fig.~3 also demonstrates that as $x_{0}$ increases, the staircase
$\mu(\zeta)$ dependence with well-defined steps and plateaus evolves
into an increasing function with modulations: the {}``plateaus''
develop noticeable slope and the {}``steps'' acquire a width. Moreover
for sufficiently large $x_{0}\gtrsim1$ the staircase-like appearance
seems to emerge only for a sufficiently deep well, i. e. large $\zeta$.
Another feature is the presence of a negative offset which is a
growing function of $x_{0}$. This is due to the fact that for $U_{s}(r)=0$
the quantum defect is a monotonically decreasing negative function of
$x_{0}$ as shown in Fig.~1.

\subsection{Zel'dovich effect in the $r_{0}\gg a_{B}$ limit}

In the $x_{0}\gg1$ limit Eq.(\ref{qdgeneral}) simplifies to the form
\begin{equation}
{\frac{a_{s}}{r_{0}}}={\frac{2}{x_{0}}}\cot(\pi\mu+x_{0}-{\frac{3\pi}{4}})+1\label{qdatomiclimit}\end{equation}
 allowing model-independent treatment. The analysis of Eq.(\ref{qdatomiclimit})
is convenient to conduct in terms of the reduced quantum defect $M=\mu+x_{0}/\pi-3/4$
whose zero gives the $x_{0}\gg1$ asymptotic of $\mu$ for $U_{s}(r)=0$.
Eq.(\ref{qdatomiclimit}) can be investigated in a manner analogous
to that of Eq.(\ref{qdzel'dovichlimit}); a brief summary is given
below. 
\begin{figure}
\includegraphics[
 width=1.0\columnwidth,
 keepaspectratio]{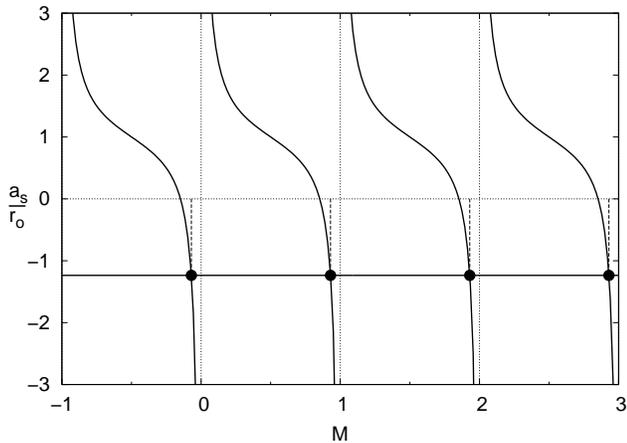}

\caption{Graphical solution of Eq.(\ref{qdatomiclimit}); $x_{0}=4$ has been
used to construct the graph. The reduced quantum defect $M=\mu+x_{0}/\pi-3/4$
is given by the intersections of the right-hand side of (\ref{qdatomiclimit})
with the line of constant $a_{s}/r_{0}$.}

\end{figure}

Fig.~4 shows the dependence of $a_{s}/r_{0}$ on $M=\mu+x_{0}/\pi-3/4$
given by Eq.(\ref{qdatomiclimit}); its inverse $M(a_{s}/r_{0})$
is a multivalued function consisting of a series of decreasing step-like
segments sandwiched between nearest integers.

The magnitude of the slope of $M(a_{s}/r_{0})$ is smallest at integer
$M$ which occurs at binding resonances, $a_{s}=\pm\infty$, i. e.
when the antinode of the function $\chi$ in Eq.(\ref{SE}) coincides
with the boundary of the inner region $r_{0}$. In the vicinity of
integer $M$ we find $M=2r_{0}/\pi x_{0}a_{s}~(mod~1)$. This translates
into an explicit result for the quantum defect $\mu=3/4-x_{0}/\pi+2r_{0}/\pi x_{0}a_{s}~(mod~1)$
valid in the limit $x_{0}\gg1$ and $r_{0}/x_{0}a_{s}\ll1$, thus
roughly covering the range of $|a_{s}|$ from $r_{0}$ to infinity.
In the vicinity of the first binding resonance we have $\mu=3/4-x_{0}/\pi+2r_{0}/\pi x_{0}a_{s}$
which is represented by the leftmost intersection in Fig.~4.

The magnitude of the slope of the $M(a_{s}/r_{0})$ dependence is
largest at half-integer $M$ which occurs at $a_{s}=r_{0}$, i. e.
when the node of the function $\chi$ in Eq.(\ref{SE}) coincides
with the boundary of the inner region $r_{0}$. Since the reduced
quantum defect $M$ is a decreasing function of $a_{s}/r_{0}$, and
the scattering length $a_{s}$ is a decreasing function of the well
depth $w$ \cite{Popov2}, then for fixed $x_{0}$ the parameter $M$
(and thus the original quantum defect $\mu$) is an increasing function
of $w$.

In contrast to the $x_{0}\ll1$ regime, here the step-plateau features
of the function $M(a_{s}/r_{0})$ are generally suppressed in the
$M(w)$ dependence. This is because the dependence of the scattering
length $a_{s}$ on the depth of the well $w$ is weakest in the region
$a_{s}\simeq r_{0}$ where the $M(a_{s}/r_{0})$ dependence shows
a {}``step''. By the same token the ``plateaus'' acquire a noticeable
slope since the $a_{s}(w)$ dependence is strongest near the binding
resonance, $a_{s}=\pm\infty$, i. e. where the $M(a_{s}/r_{0})$ dependence
is weakest. As a result the dependence of the quantum defect $\mu$
on the depth of the well $w$ is more appropriately viewed as consisting
of modulations superimposed on an increasing curve. These modulations
still have their origin in the binding properties of the inner potential
$U_{s}(r)$. 
\begin{figure}
\includegraphics[
 width=1.0\columnwidth,
 keepaspectratio]{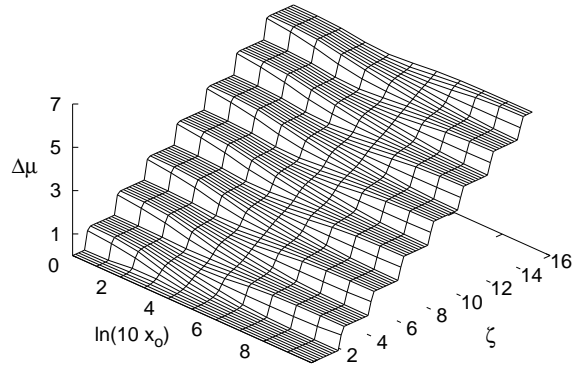} 

\caption{Plot of the surface of the relative quantum defect $\Delta\mu(\zeta,x_{0})=\mu(\zeta,x_{0})-\mu(0,x_{0})$
for a rectangular potential well with dimensionless range and depth
parameters $x_{0}$ and $\zeta$, respectively, according to Eq.(\ref{qdgeneral}).}

\end{figure}

\subsection{Rectangular well example}

The analysis of Sections IIA and IIC is illustrated in Fig.~5 where
using the example of the rectangular well and Eq.(\ref{qdgeneral})
we plot the surface of the relative quantum defect $\Delta\mu(\zeta,x_{0})=\mu(\zeta,x_{0})-\mu(0,x_{0})$.
The point of subtracting $\mu(0,x_{0})$ from $\mu(\zeta,x_{0})$
is to isolate the physics of binding from the background $\mu(0,x_{0})$
which is a monotonically decreasing function of $x_{0}$ shown in Fig.~1.
The peculiar shape of the resulting surface can then be understood
as follows:

For $x_{0}\ll1$ the background contribution $\mu(0,x_{0})$ is negligible
(see Fig.~1), the relative quantum defect $\Delta\mu(\zeta,x_{0})$
reduces to $\mu(\zeta,x_{0})$ which, according to our earlier analysis,
is a staircase function of $\zeta$ with steps located at odd $\zeta$,
i. e. when the bound states occur in the well.

For $x_{0}\gg1$ the background contribution $\mu(0,x_{0})$ is $3/4-x_{0}/\pi$,
and we find that $\Delta\mu(\zeta,x_{0})=\mu(\zeta,x_{0})-\mu(0,x_{0})=\mu(\zeta,x_{0})+x_{0}/\pi-3/4=M$
which, according to Fig.~4, is a decreasing staircase function of
$a_{s}/r_{0}$. For not very deep well the dependence on $\zeta$
has a form of a rounded staircase with {}``steps'' and {}``plateaus''
centered at even ($a_{s}=r_{0}$) and odd ($a_{s}=\pm\infty$) values
of $\zeta$, respectively. In this regime the underlying step-plateau
character of the $M(a_{s}/r_{0})$ function is preserved in the $M(\zeta)$
dependence due to the appreciable slope of the $a_{s}(\zeta)$ dependence
at $a_{s}=r_{0}$ and relatively weak divergence at $a_{s}=\pm\infty$.
To recapitulate, both for $x_{0}\ll1$ and $x_{0}\gg1$ and not very
deep well the relative quantum defect $\Delta\mu(\zeta,x_{0})$ is
an increasing staircase function of the depth parameter $\zeta$ with
the steps located at odd ($x_{0}\ll1$) or even ($x_{0}\gg1$) values
of $\zeta$. The crossover between the two regimes can be seen in
Fig.~5 as a relatively narrow stripe of very weak modulations.

A qualitatively different staircase-like dependence emerges for $x_{0}\gg1$
and sufficiently deep well because as $\zeta\rightarrow\infty$ the
slope of the $a_{s}(\zeta)$ function at $a_{s}=r_{0}$ tends to zero
while the binding resonances, $a_{s}=\pm\infty$, become progressively
more singular. As a result the {}``steps'' and {}``plateaus''
switch places - the former becomes centered at odd while the latter
at even values of $\zeta$. This is somewhat similar to what happens
in the $x_{0}\ll1$ limit. This observation explains why the small
modulation crossover stripe in Fig.~5 runs at an angle to the $(\Delta\mu,\zeta)$
plane. The important qualitative difference between the $x_{0}\ll1$
and $x_{0},\zeta\gg1$ staircases is that the latter have {}``plateaus''
centered at half-integer values of $M$, thus corresponding to the
coincidence of a node of the function $\chi$ in Eq.(\ref{SE}) with
the boundary of the inner region $r=r_{0}$.

\subsection{Semiclassical treatment}

For most realistic models of the inner potential $U_{s}(r)$ the exact
analytical calculation of the scattering length $a_{s}$ entering
the general expression for the quantum defect Eq.(\ref{qdgeneral})
may not be possible. Therefore it is pertinent to understand whether
there is an approximate analytical treatment capturing the Zel'dovich
spectrum reconstruction. This is especially relevant to the $x_{0}\gg1$
regime when the phenomenon manifests itself only as a modulation of
the quantum defect superimposed on a monotonic curve.

For $x_{0}\gg1$ and sufficiently smooth $U_{s}(r)$ the standard
semiclassical approximation is applicable, and the corresponding solution
to Eq.(\ref{zeroSE}) can be written as \begin{equation}
\chi_{sc}(r)\sim\left(S'(r)\right)^{-1/2}\sin{\frac{S(r)}{\hbar}},\label{scstandard}\end{equation}
where \begin{equation}
S(r)=\int\limits _{0}^{r}\left(-2mU_{s}(r)\right)^{1/2}dr\label{action}\end{equation}
is the classical action acquired by a zero-energy particle moving
radially out from zero to $r$, and the prime in Eq.(\ref{scstandard})
denotes differentiation with respect to $r$. The semiclassical expression
for the scattering length which can be deduced from Eq. (\ref{scstandard})
with the help of Eq. (\ref{logderscatteringlength}) has been given
by Berry \cite{Berry}\begin{equation}
{\frac{a_{s}}{r_{0}}}=1-\left(-{\frac{\hbar^{2}}{2mr_{0}^{2}U_{s}(r_{0})}}\right)^{1/2}\tan{\frac{S_{0}}{\hbar}}\label{scscatteringlength}\end{equation}
 where $S_{0}\equiv S(r_{0})$. Eqs.(\ref{scstandard}) and (\ref{scscatteringlength}),
generalizing the {}``rectangular well'' expressions for the wave
function and the scattering length, are applicable when the number
of de Broglie's half-waves $S_{0}/\pi\hbar$ fitting inside $U_{s}(r)$
is very large. If we additionally assume the continuity of the central
potential $U(r)$ in Eq.(\ref{SE}) at the boundary of the inner region,
\begin{equation}
U_{s}(r_{0})=-\hbar^{2}/ma_{B}r_{0},\label{continuity}\end{equation}
 then Eq.(\ref{scscatteringlength}) simplifies to $a_{s}/r_{0}=1-(2/x_{0})\tan(S_{0}/\hbar)$.
Combining this with Eq.(\ref{qdatomiclimit}) we find an explicit
semiclassical expression for the quantum defect \begin{equation}
\mu_{sc}={\frac{3}{4}}-{\frac{x_{0}}{\pi}}+{\frac{S_{0}}{\pi\hbar}}-{\frac{1}{2}}\label{scqd}\end{equation}
 which can be interpreted as approximately the sum of $3/4-x_{0}/\pi$,
the quantum defect for $U_{s}(r)=0$, and the number of de Broglie's
half-waves $S_{0}/\pi\hbar$ fitting inside the inner part of the
potential; the estimate $\mu_{sc}\simeq S_{0}/\pi\hbar$ has been
given earlier \cite{Friedrich}. The number of de Broglie's half-waves
can be estimated in terms of the dimensionless depth of the inner
well $w\simeq mr_{0}^{2}|U_{s}|/\hbar^{2}$ as $S_{0}/\pi\hbar\simeq w^{1/2}$
which implies that for fixed $x_{0}$ the quantum defect (\ref{scqd})
is a monotonically increasing function of $w$ without any modulations.
We conclude that the Zel'dovich modulations of the quantum defect
are lost in the semiclassical approximation despite the fact that
the corresponding scattering length (\ref{scscatteringlength}) does
exhibit binding resonances. Thus for $x_{0}\gg1$ a treatment better
than semiclassical is required to capture the deviations from monotonic
behavior; a similar conclusion has been reached earlier \cite{Fano}.

For the rectangular well of radius $r_{0}$ and depth $U_{0}$ characterized
by the coupling constant $\zeta=(8mU_{0}r_{0}^{2}/\pi^{2}\hbar^{2})^{1/2}$
the expression for the scattering length (\ref{scscatteringlength})
is exact, and then Eq.(\ref{scqd}) predicts that $\mu=1/4-x_{0}/\pi+\zeta/2=1/4-\zeta/2<0$.
This is the $x_{0},\zeta\gg1$ value of the quantum defect in the
middle of the small modulation stripe in Fig.~5 whose locus, $x_{0}=\pi\zeta$,
can be deduced from Eq.(\ref{continuity}). The quantum defect is
negative because for continuous $U(r)$ the short-distance rectangular
well potential is always less attractive than the Coulomb potential.

\section{Quantum defect of Rydberg electron}

Now when we understand the manifestations of the Zel'dovich effect,
and what kind of accuracy is required to approximately capture it,
we begin computing the quantum defect of the Rydberg electron as a
function of position along the Periodic Table. The quantum defect
is given by the exact result Eq.(\ref{qdgeneral}) with $r_{0}$ and
$a_{s}$, being the size and the scattering length of the residual
atomic ion, respectively, both dependent upon atomic number $Z$.
The resulting $\mu(Z)$ dependence will exhibit modulations both due
to systematic (Zel'dovich) and shell effects. As discussed in the
Introduction, the shell effects obscure systematic trends making it
difficult to see that some modulations of $\mu(Z)$ have their origin
in the binding properties of the ionic core. To circumvent this inconvenience
below we conduct a calculation capturing only systematic effects.
The comparison of the results with both experimental and numerical
data (additionally containing the shell effects) will allow us to
disentangle physically different sources of deviation from purely
monotonic behavior.

\subsection{Method of comparison equations}

The short-distance potential $U_{s}(r)$ characterizing the residual
atomic ion will be assumed to match at its boundary the Coulomb potential
of unit charge \cite{Latter},\cite{Fano} (see Eq.(\ref{continuity})).
As the Rydberg electron moves inside the ionic core, the screening
of the nuclear charge by the inner shell electrons diminishes which
implies that for $r<r_{0}$ the short-distance potential $U_{s}(r)$
is more attractive than the Coulomb potential of unit charge. Therefore
the quantum defect is a necessarily positive and increasing function
of atomic number $Z$. As $r\rightarrow0$, the inner potential approaches
that of a nucleus of charge $Ze$, i. e. $U_{s}(r\rightarrow0)\rightarrow-Ze^{2}/r=-Z\hbar^{2}/ma_{B}r$,
and Eq.(\ref{zeroSE}) reduces to \begin{equation}
{\frac{d^{2}{\chi}}{dr^{2}}}+{\frac{2Z}{ra_{B}}}\chi=0\label{nearnucleousSE}\end{equation}
 This presents a convenient starting point for obtaining an approximate
solution to the differential equation (\ref{zeroSE}) via the method
of comparison equations as described by Berry and Mount \cite{BM}.
Since Eq.(\ref{nearnucleousSE}) is exactly solvable, and the potentials
of Eqs.(\ref{zeroSE}) and (\ref{nearnucleousSE}) are somewhat similar,
the solution to (\ref{zeroSE}) should be also similar to that of
(\ref{nearnucleousSE}) and can be transformed into it by a slight deformation
of coordinates and an amplitude adjustment. The details of finding
an appropriate mapping are given in Ref. \cite{BM}; the resulting
approximate solution of (\ref{zeroSE}) is then given by \begin{equation}
\chi(r)\sim\left({\frac{S(r)}{S'(r)}}\right)^{1/2}J_{1}\left({\frac{S(r)}{\hbar}}\right)\label{BMfunction}\end{equation}
 The method of comparison equations includes the conventional semiclassical
treatment as a special case \cite{BM}. From this more general viewpoint
Eq. (\ref{scstandard}) can be viewed as a result of the deformation
and amplitude adjustment of the {}``rectangular well'' sine solution.

To assess the accuracy of (\ref{BMfunction}) let us first look at
the limit $S(r)/\hbar\ll1$. According to Eq.(\ref{action}) this
corresponds to $r\rightarrow0$ when $U_{s}(r)\rightarrow-Z\hbar^{2}/ma_{B}r$.
Then $S(r\rightarrow0)/\hbar\rightarrow(8Zr/a_{B})^{1/2}$ and $\chi(r\rightarrow0)\sim r^{1/2}J_{1}(\sqrt{8Zr/a_{B}})$
which can be recognized as the solution to (\ref{nearnucleousSE}).

In the opposite limit $S(r)/\hbar\gg1$ a semiclassical approximation
is expected to be valid and Eq.(\ref{BMfunction}) simplifies to $\chi(r)\sim\left(S'(r)\right)^{-1/2}\sin(S(r)/\hbar-\pi/4)$.
This is similar to the naive semiclassical result (\ref{scstandard})
with the extra phase of $-\pi/4$ correcting for the failure of the
standard semiclassical treatment in the Coulomb field of charge $Ze$
at distances $r\lesssim a_{B}/Z$. Thus for the inner potential $U_{s}(r)$
which has a Coulombic singularity as $r\rightarrow0$ but is otherwise
smooth the analog of Eq.(\ref{scqd}) is \begin{equation}
\mu_{sc}=-{\frac{x_{0}}{\pi}}+{\frac{S_{0}}{\pi\hbar}}\label{Coulombscqd}\end{equation}

The expression for the scattering length corresponding to Eq.(\ref{BMfunction})
can be found with the help of Eq.(\ref{logderscatteringlength}) \begin{equation}
{\frac{a_{s}}{r_{0}}}=1-4\left(x_{0}\left({\frac{\hbar}{S_{0}}}+{\frac{J_{0}({\frac{S_{0}}{\hbar}})-J_{2}({\frac{S_{0}}{\hbar}})}{J_{1}({\frac{S_{0}}{\hbar}})}}\right)-{\frac{r_{0}U_{s}'(r_{0})}{U_{s}(r_{0})}}\right)^{-1}\label{BMscatteringlength}\end{equation}
 where we also used the condition of continuity (\ref{continuity}).
Eq.(\ref{BMscatteringlength}) can be used to go beyond the semiclassical
expression (\ref{Coulombscqd}). Combining Eqs.(\ref{qdatomiclimit})
and (\ref{BMscatteringlength}) we find that in the $x_{0},S_{0}/\hbar\gg1$
limit the quantum defect can be presented as $\mu=\mu_{sc}+\delta\mu$
where $\mu_{sc}$ is the semiclassical answer (\ref{Coulombscqd})
and the correction, \begin{equation}
\delta\mu={\frac{r_{0}U_{s}'(r_{0})}{4\pi x_{0}U_{s}(r_{0})}}\left(1-\sin{\frac{2S_{0}}{\hbar}}\right),\label{zel'oscillation}\end{equation}
 captures the Zel'dovich effect now manifesting itself as a simple harmonic
modulation superimposed on the semiclassical background. The period
of the oscillation is exactly one de Broglie's half-wave while the
amplitude is of the order $x_{0}^{-1}$. The fact that the latter
is independent of the number of de Broglie's half-waves fitting inside
$U_{s}(r)$ implies that the Zel'dovich effect persists for any value
of $S_{0}/\pi\hbar$.

\subsection{Thomas-Fermi model of atomic ion: semiclassical solution}

Below we follow Latter \cite{Latter} and assume that the potential
of the atomic ion $U_{s}(r)$ can be approximated by the Thomas-Fermi
theory \cite{TF}: \begin{equation}
U_{s}(r)=-{\frac{\hbar^{2}Z}{ma_{B}r}}\phi({\frac{rZ^{1/3}}{ba_{B}}})\label{TFpotential}\end{equation}
 where $b=(3\pi/4)^{1/2}/2\simeq0.885$, and the universal function
$\phi(y)$ is the solution to the nonlinear Thomas-Fermi equation
\begin{equation}
y^{1/2}{\frac{d^{2}\phi}{dy^{2}}}=\phi^{3/2}\label{TFfunction}\end{equation}
 subject to the boundary conditions $\phi(0)=1$ and $\phi(\infty)=0$
\cite{TF}. Then the size of the ion $r_{0}$ and thus the range parameter
$x_{0}$ (\ref{x0}) are determined by the continuity condition (\ref{continuity}),
i. e. when the Thomas-Fermi potential (\ref{TFpotential}) meets the
Coulomb potential of unit charge: \begin{equation}
\phi({\frac{x_{0}^{2}Z^{1/3}}{8b}})={\frac{1}{Z}}\label{TFrange}\end{equation}
 Eqs. (\ref{TFpotential}) and (\ref{TFrange}) imply that the natural
variable to characterize the strength of the potential of the atomic
ion is $Z^{1/3}$. Indeed, the typical length scale of the Thomas-Fermi
theory is $a_{B}/Z^{1/3}$, the magnitude of the typical potential
is $(Z\hbar^{2}/ma_{B})(Z^{1/3}/a_{B})=(\hbar^{2}/ma_{B}^{2})Z^{4/3}$,
and thus the dimensionless coupling constant $w\simeq mr_{0}^{2}|U_{s}|/\hbar^{2}$
which entered the general analysis of Section II is of the order $Z^{2/3}$;
the parameter $Z^{1/3}$ then parallels $\zeta\simeq w^{1/2}$ used
in the rectangular well example of the inner potential.

Since the Thomas-Fermi function $\phi(y)$ is a monotonically decreasing
function of its argument, the $x_{0}(Z^{1/3}$) dependence defined
through Eq.(\ref{TFrange}) is a monotonically increasing function
of $Z^{1/3}$. The boundary condition $\phi(0)=1$ implies that $x_{0}(1)=0$
which is in accordance with the expectation that for Hydrogen ($Z=1$)
we have the standard Coulomb problem in the whole space ($x_{0}=0$).
As evident from (\ref{TFrange}) small values of $\phi$ are relevant
for large $Z$; in view of $\phi(y\rightarrow\infty)\rightarrow144/y^{3}$
\cite{TF} this means that there is an upper bound to the range parameter,
$x_{0}(\infty)=2^{13/6}3^{1/3}b^{1/2}\simeq6.092$. 

Another quantity of interest is the number of de Broglie's half-waves
fitting inside the Thomas-Fermi atomic ion 
\begin{eqnarray}
\label{TFaction}
{\frac{S_{0}}{\pi\hbar}} & = & \int\limits _{0}^{r_{0}}\left({\frac{-2mU_{s}(r)}{\,\,\pi^{2}\hbar^{2}}}\right)^{1/2}dr\nonumber \\
 & = & {\frac{(2b)^{1/2}Z^{1/3}}{\pi}}\int\limits _{0}^{x_{0}^{2}Z^{1/3}/8b}\left({\frac{\phi(y)}{y}}\right)^{1/2}dy\end{eqnarray}
 which is a monotonically increasing function of $Z^{1/3}$: for $Z\rightarrow1$
it vanishes as $x_{0}/\pi$ while for large $Z$ we have $S_{0}/\pi\hbar\sim Z^{1/3}$,
a known result \cite{Migdal}.

For intermediate values of the atomic number the $Z^{1/3}$ dependences
of the range parameter $x_{0}$ and the number of de Broglie's half-waves
$S_{0}/\pi\hbar$ can be found by numerically solving the Thomas-Fermi
equation (\ref{TFfunction}), inverting (\ref{TFrange}), and computing
the integral (\ref{TFaction}). The results are displayed in Fig.~6
where we show $x_{0}/\pi$ and $S_{0}/\pi\hbar$ as functions of $Z^{1/3}$.
These functions are used to also plot the semiclassical quantum defect
given by Eq.(\ref{Coulombscqd}). To assess the accuracy of the resulting
$\mu_{sc}(Z^{1/3})$ dependence we need to verify whether approximations
used to derive Eq.(\ref{Coulombscqd}) are adequate. The first assumption,
$x_{0}\gg1$, is equivalent to the assertion that the quantum defect
for $U_{s}(r)=0$ can be replaced by its large $x_{0}$ limit. Looking
at Fig.~1 where the dependences in question are compared we conclude
that {}``large'' here really means $x_{0}\gtrsim\pi$. The second,
semiclassical assumption $S_{0}/\hbar\gg1$ in practice has a good
accuracy provided $S_{0}/\pi\hbar$, the number of de Broglie's half-waves
fitting into the inner potential $U_{s}(r)$, is anything more than
one or two. Inspecting Fig.~6 we see that the conditions $x_{0}/\pi\gtrsim1$
and $S_{0}/\pi\hbar\gtrsim2$ are satisfied for $Z\gtrsim8$. This is
also the practical condition for the Thomas-Fermi-Latter model of the
residual ion to be applicable.  We note
additionally that although Eq.(\ref{Coulombscqd}) is not expected
to be valid for smallest $Z$, the limit $\mu(1)=0$ is nevertheless
correctly reproduced and for any $Z$ the semiclassical quantum defect
is an increasing positive function of $Z^{1/3}$ in accordance with
physical expectation. We conclude that except possibly for the elements
of the first row of the Periodic Table, the semiclassical result for
the quantum defect Eq.(\ref{Coulombscqd}) shown in Fig.~6 is accurate;
only qualitative agreement is expected for lightest elements. 
\begin{figure}
\includegraphics[
 width=1.0\columnwidth,
 keepaspectratio]{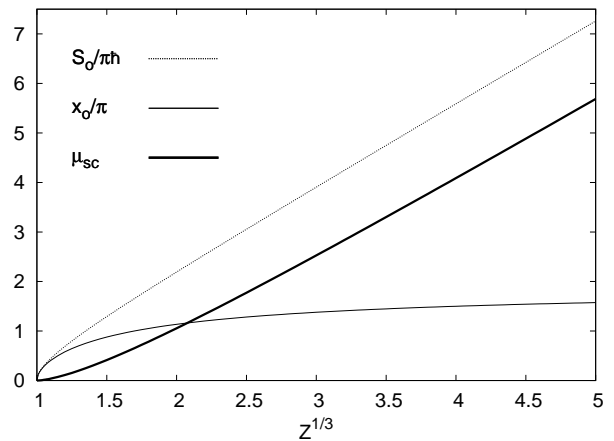} 

\caption{Plots of the number of de Broglie's half-waves $S_{0}/\pi\hbar$,
(\ref{TFaction}), the range parameter $x_{0}/\pi$, (\ref{TFrange}),
and semiclassical quantum defect $\mu_{sc}$, (\ref{Coulombscqd})
as functions of $Z^{1/3}$ for the Thomas-Fermi model of the residual
atomic ion. }
\end{figure}

\subsection{Beyond semiclassical approximation: connection to binding
properties of ionic core}

A more accurate $\mu(Z^{1/3})$ dependence can be found by computing
the scattering length (\ref{BMscatteringlength}) and substituting
the outcome together with the $x_{0}(Z^{1/3})$ dependence, Fig.~6,
in our general expression for the quantum defect (\ref{qdgeneral}).
The result is shown in Fig.~7 where we also plot the semiclassical quantum
defect, $\mu_{sc}$, Eq.(\ref{Coulombscqd}) (gray scale). It now
becomes obvious that for any $Z$ the bulk contribution into the quantum
defect is well-captured semiclassically. The Zel'dovich spectral modulation
clearly visible in Fig.~7 is a relatively weak effect. To separate
the modulation from the monotonic s semiclassical background the inset
shows the difference $\delta\mu=\mu-\mu_{sc}$ which appears to be
a nearly periodic function of $Z^{1/3}$. To better understand the meaning
of this periodicity the inset also shows the limiting expression (\ref{zel'oscillation})
(gray scale) \cite{note2}. Both curves are exactly in phase and for
sufficiently heavy elements their magnitudes agree semi-quantitatively.
This observation implies that the Zel'dovich modulation is a periodic
function of the number of de Broglie's half-waves fitting inside the
ionic core of the atom. 
\begin{figure}
\includegraphics[
 width=1.0\columnwidth,
 keepaspectratio]{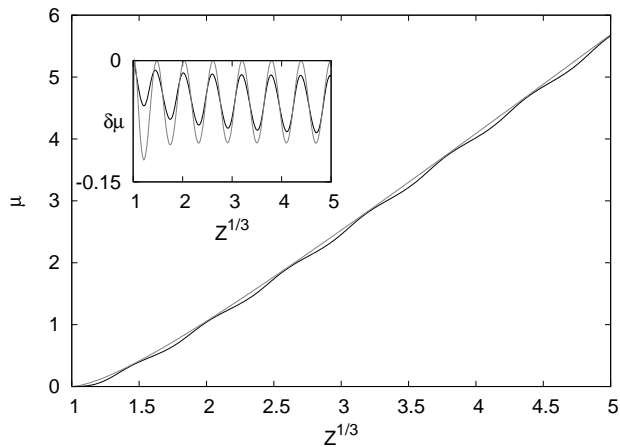} 

\caption{Dependence of the quantum defect $\mu$ on $Z^{1/3}$ along with
its semiclassical approximant $\mu_{sc}$, Eq.(\ref{Coulombscqd})
(gray scale). The inset shows the Zel'dovich modulation $\delta\mu(Z^{1/3})=\mu-\mu_{sc}$
together with the limiting expression, Eq.(\ref{zel'oscillation})
(gray scale).}
\end{figure}

A complementary way to see the connection between the spectral modulation
and the binding properties is presented in Fig.~8 where we compare
the $\delta\mu(Z^{1/3})$ dependence with the behavior of the scattering
length of the residual atomic ion (\ref{BMscatteringlength}). The
latter, numerically computed for the Thomas-Fermi model of the residual
atomic ion, Eqs.(\ref{TFpotential}) - (\ref{TFaction}), is shown
in gray scale. The binding singularities of the scattering length
are nearly equidistant confirming the earlier observation that the parameter
$Z^{1/3}$ is analogous to $\zeta$ used in Figs.~3 and 5 to display
the Zel'dovich effect for the rectangular well model of the inner
potential. A qualitatively similar behavior of the scattering length
of the Thomas-Fermi atom as a function of $Z$ has been reported by
Robinson \cite{Robinson}; quantitative differences may be attributed
to the assumption \cite{Robinson} that the Thomas-Fermi potential
vanishes at a distance of the order $a_{B}/Z^{1/3}$ which is different
from our choice of $r_{0}$, Eqs.(\ref{continuity}) and (\ref{TFrange}).

Fig.~8 makes it clear that the maxima of the oscillation $\delta\mu$
occur when $a_{s}=r_{0}$, i. e. when the node of the function $\chi(r)$
in Eq.(\ref{SE}) coincides with the boundary of the atomic ion. On
the other hand, the minima of $\delta\mu$ are correlated with binding
singularities of the ionic core, $a_{s}=\pm\infty$, thus corresponding
to the antinode of $\chi(r)$ being near the ion boundary.

\begin{figure}
\includegraphics[ width=1.0\columnwidth,
 keepaspectratio]{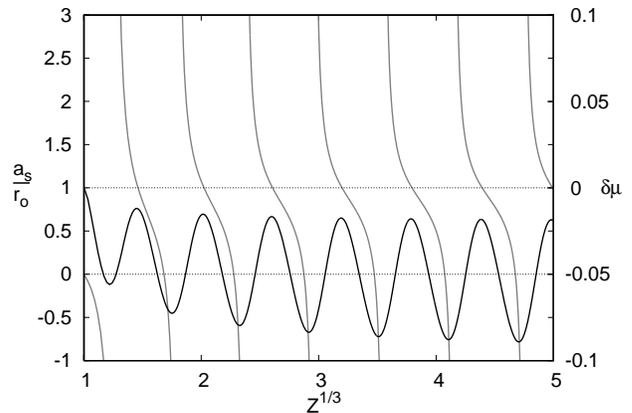}

\caption{Dependences of the Zel'dovich modulation $\delta\mu=\mu-\mu_{sc}$
and the reduced scattering length of the ionic core $a_{s}/r_{0}$
(gray scale) on $Z^{1/3}$. The lines $a_{s}/r_{0}=0$ and $a_{s}/r_{0}=1$
are also shown to help the eye.}

\end{figure}

\section{Comparison with experimental and numerical data}

\squeezetable
\begingroup
\begin{table}

\caption{Experimentally measured quantum defects for series of elements with
their atomic numbers $Z$ and corresponding references. Systematic
quantum defects of this work are also displayed for comparison.}

\begin{tabular}{|c|c|c|c|c|}
\hline 
Z&
 Element&
 Experimental $\mu$&
 Reference&
 Systematic $\mu$\tabularnewline
\hline
2&
 He&
 .139&
 \cite{RefData}&
 .110\tabularnewline
\hline
3&
 Li&
 .400&
 \cite{alkali}&
 .336\tabularnewline
\hline
4&
 Be&
 .670&
 \cite{TopBase}&
 .478\tabularnewline
\hline
5&
 B&
 1.000&
 \cite{TopBase}&
 .600\tabularnewline
\hline
6&
 C&
 1.050&
 \cite{TopBase}&
 .744\tabularnewline
\hline
7&
 N&
 1.091&
 \cite{TopBase}&
 .904\tabularnewline
\hline
8&
 O&
 1.132&
 \cite{TopBase}&
 1.040\tabularnewline
\hline
9&
 F&
 1.203&
 \cite{TopBase}&
 1.144\tabularnewline
\hline
10&
 Ne&
 1.300&
 \cite{noble}&
 1.229\tabularnewline
\hline
11&
 Na&
 1.348&
 \cite{alkali}&
 1.307\tabularnewline
\hline
12&
 Mg&
 1.517&
 \cite{TopBase}&
 1.388\tabularnewline
\hline
13&
 Al&
 1.758&
 \cite{Al}&
 1.476\tabularnewline
\hline
14&
 Si&
 1.816&
 \cite{TopBase}&
 1.574\tabularnewline
\hline
16&
 S&
 1.947&
 \cite{TopBase}&
 1.774\tabularnewline
\hline
17&
 Cl&
 2.128&
 \cite{Cl}&
 1.861\tabularnewline
\hline
18&
 Ar&
 2.140&
 \cite{noble}&
 1.935\tabularnewline
\hline
19&
 K&
 2.180&
 \cite{alkali}&
 1.999\tabularnewline
\hline
20&
 Ca&
 2.340&
 \cite{RefData}&
 2.056\tabularnewline
\hline
22&
 Ti&
 2.400&
 \cite{TiVFe}&
 2.161\tabularnewline
\hline
23&
 V&
 2.300&
 \cite{TiVFe}&
 2.134\tabularnewline
\hline
26&
 Fe&
 2.600&
 \cite{TiVFe}&
 2.390\tabularnewline
\hline
29&
 Cu&
 2.600&
 \cite{RefData}&
 2.594\tabularnewline
\hline
30&
 Zn&
 2.639&
 \cite{Zn}&
 2.660\tabularnewline
\hline
36&
 Kr&
 3.100&
 \cite{noble}&
 2.956\tabularnewline
\hline
37&
 Rb&
 3.131&
 \cite{rubidium}&
 2.994\tabularnewline
\hline
38&
 Sr&
 3.269&
 \cite{RefData}&
 3.031\tabularnewline
\hline
39&
 Y&
 3.385&
 \cite{PtY}&
 3.067\tabularnewline
\hline
42&
 Mo&
 3.476&
 \cite{Mo}&
 3.180\tabularnewline
\hline
47&
 Ag&
 3.600&
 \cite{silver}&
 3.404\tabularnewline
\hline
49&
 In&
 3.720&
 \cite{In}&
 3.503\tabularnewline
\hline
54&
 Xe&
 4.000&
 \cite{noble}&
 3.722\tabularnewline
\hline
55&
 Cs&
 4.049&
 \cite{cesium}&
 3.759\tabularnewline
\hline
56&
 Ba&
 4.200&
 \cite{RefData}&
 3.793\tabularnewline
\hline
70&
 Yb&
 4.280&
 \cite{RefData}&
 4.193\tabularnewline
\hline
78&
 Pt&
 4.611&
 \cite{PtY}&
 4.479\tabularnewline
\hline
79&
 Au&
 4.660&
 \cite{gold}&
 4.515\tabularnewline
\hline
83&
 Bi&
 4.890&
 \cite{RefData}&
 4.645 \tabularnewline
\hline
\end{tabular}
\end{table}
\endgroup

We found experimental values of quantum defects for 37 elements of the
Periodic Table.  These data and their sources are compiled in Table 1
where we also list systematic quantum defects of our work (also
displayed in Fig.~7).  Some of the figures which we regard as
``experimental'' came from an on-line database \cite{TopBase} where the 
quantum defect is computed from available spectroscopic data.  

In cases of He, Be, Mg, Ca, and Mo there is more than one value of the
quantum defect available depending on the angular momentum of the
ionic core of the atom.  Since our theory represents average
properties and does not distinguish between different $LS$ terms of an
atomic configuration,  in Table 1 we chose to show only the values 
corresponding to lowest angular momentum of the ionic core.  It turns
out they better agree with our calculation than those left out.

The works of Manson \cite{Manson} and Fano, Theodosiou and Dehmer
\cite{Fano} contain graphs of numerically evaluated $\mu(Z)$
dependence for all elements.  After verifying that the results of
both studies are nearly identical, we chose to restrict ourselves to
those of the later Ref.\cite{Fano}.  

\begin{figure}
\includegraphics[ width=1.0\columnwidth,
 keepaspectratio]{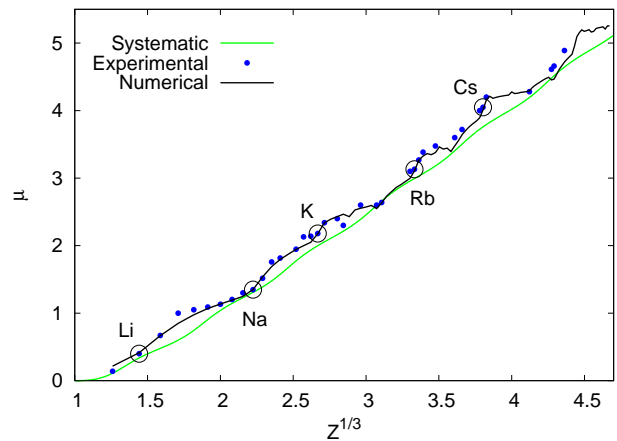}

\caption{Systematic, experimental, and numerical dependences of the
quantum defect $\mu$ on $Z^{1/3}$.  To help orientation within the
Periodic Table experimental alkali data are circled.}

\end{figure}
  
Experimental, numerical and systematic $\mu(Z^{1/3})$ dependences are
displayed in Fig.~9.  In order to produce the numerical curve, the
data \cite{Fano} have been scanned, digitized, and replotted as
a function of $Z^{1/3}$.  We also circled locations of alkali metals 
because their ionic cores have noble element electronic configurations
thus marking (for the Rydberg atom problem) the end of a period.  

Fig.~9 makes it clear that all three dependences are in fairly good
agreement and $Z^{1/3}$ is certainly the right variable to use for analysis. 
It is not surprising that numerical results \cite{Fano} are generally in better
agreement with experimental data than our systematic findings because
our calculation omits the shell effects.  

\begin{figure}
\includegraphics[ width=1.0\columnwidth,
keepaspectratio]{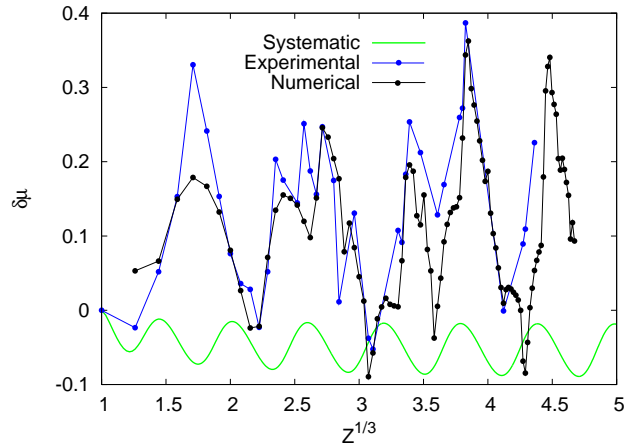}

\caption{Modulation of systematic, experimental and numerical
quantum defect $\delta\mu = \mu - \mu_{sc}$ relative to the
semiclassical background, Eq.(\ref{Coulombscqd}) as a function of
$Z^{1/3}$.}

\end{figure} 

\subsection{Effects of shell structure}

In order to be able to separate systematic and shell effects, in Fig.~10
we display $\delta \mu = \mu -
\mu_{sc}$, the modulation of the quantum defect relative to the monotonic
semiclassical background, Eq.(\ref{Coulombscqd}),  which accounts for the 
bulk of the quantum defect value.  Fig.~10 shows that the experimental
and numerical variations of the quantum defect are bounded which is
consistent with the view that they are due to repetitive physics.  
Moreover, the systematic modulation due to the Zel'dovich effect
appears to have an amplitude which is several times smaller than
those of experimental and numerical data.  This observation implies
that it may be possible to understand gross features of the
experimental and numerical modulations of the quantum defect as mostly
due to the effects of the shell structure.

This viewpoint can be supported by qualitative analysis which
rests on the semiclassical result (\ref{Coulombscqd}).  First, let us
anticipate the outcome of incorporating the shell effects into the
calculation.  This amounts to replacing the smooth inner potential 
$U_{s}(r)$ by one with modulations due to the spatial 
variation of the electron density reflecting the shell
structure.  This replacement will result in a value of the
size of the ionic core $r_{0}$, Eq.(\ref{continuity}), generally 
different from its systematic counterpart.  

Let us additionally assume that the inner potential $U_{s}(r)$
with shell effects included is still sufficiently smooth so that
a semiclassical treatment is valid. The corresponding quantum defect
(\ref{Coulombscqd}) will deviate away from the systematic result due to
different values of the range parameter $x_{0}$, Eq.(\ref{x0}), and 
the number of de Broglie's half-waves, $S_{0}/\pi \hbar$.  Because the
latter involves the integral of $(-U_{s}(r))^{1/2}$ from zero to
$r_{0}$ (see Eqs.(\ref{action}) and (\ref{TFaction})) the modulations
above and below the systematics present in $U_{s}(r)$ are expected to largely
cancel each other and the deviation from our results can be mostly
attributed to the different size of the ionic core.

This argument implies that the quantum defect is strongly sensitive to
the value of the size of ionic core $r_{0}$ and weakly sensitive to the
details of the inner potential $U_{s}(r)$.  In reality the inner
potential may not be smooth enough for the semiclassial treatment to
be quantitatively correct.  Therefore we do not expect more than a
qualitative insight into the trends of the variations of the quantum
defect induced by the shell effects.       

The simple rule that emerges can be most easily deduced from Fig.~6 by
keeping in mind the relationship between the range parameter $x_{0}$,
Eq.(\ref{x0}) and the size of the ionic core $r_{0} \propto x_{0}^{2}$:  deviation in
$r_{0}$ away from systematics leads to the same sign deviation in
the quantum defect.
\begin{figure}
\includegraphics[ width=1.0\columnwidth,
keepaspectratio]{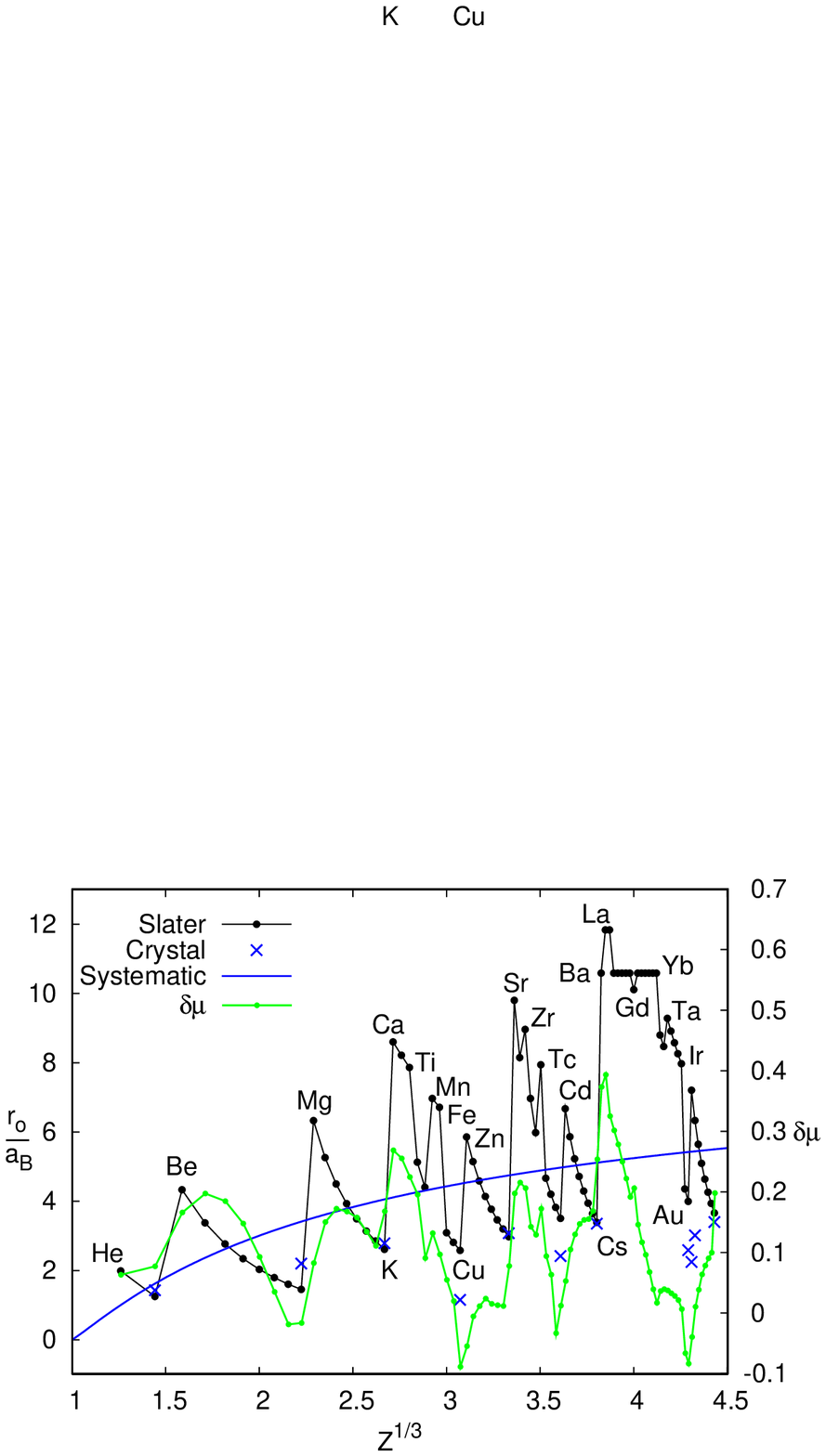}

\caption{Slater's ionic radii for singly-charged positive ions
together with a series of corresponding
ionic radii in crystals and systematic sizes of ionic core of the
Rydberg atom, all in atomic units, as functions of $Z^{1/3}$.
Numerical variation of the quantum defect $\delta \mu = \mu -
\mu_{sc}$ is also displayed to show the correlation with Slater's radii.}

\end{figure}
Since the size of the ionic core of the Rydberg atom has a physical
meaning close to that of an ionic radius, to verify the correlation we need a set of ionic radii for
singly-charged positive ions as a function of position $Z$ along the
Periodic Table.  

Seventy five years ago J. C. Slater \cite{Slater}
gave a very useful, general, empirical set of rules to
approximate analytically atomic wave functions for all the elements in 
any stage of ionization.  The radial part of the single-electron wave
function is selected in the form
\begin{equation}
\label{Slaterwf}
\psi(r) \propto r^{n^{*} - 1} e^{- Z^{*}r/n^{*}a_{B}}
\end{equation}
which can be recognized as the large-distance asympotics of a
Hydrogen-like wave function with an effective quantum number $n^{*}$ and
an effective nuclear charge $Z^{*}e$.  Based on the underlying
electronic structure, Slater's rules assign values of
$n^{*}$ and $Z^{*}$ to the electrons of each shell of an atom or ion,
so that a complete set of single-electron wave functions can be constructed.

For a given shell the maximum of the electron density 
$4\pi r^{2}\psi^{2}(r)$ is located at 
\begin{equation}
\label{rmax}
r_{max}/a_{B} = (n^{*})^{2}/Z^{*}
\end{equation} 
which formally coincides with the expression for the radius of the 
corresponding circular
orbit in Bohr's old theory.  The radius of the maximum density of the
outermost shell is expected to correlate with the size of the atom or ion.
Specifically, Slater defines an ionic radius $r_{0} > r_{max}$ as a 
distance at which the electron density becomes 10\%\ of its maximal value.    
 
Using the existing knowledge of electronic configurations \cite{NIST}
we applied Slater's rules to calculate the ionic radii of
singly-charged positive ions.  The result is shown in Fig.~11 where the 
elements marking the beginning or an end of more dramatic changes
in the ionic radius are labeled.  For comparison we also displayed
a series of ionic crystal radii \cite{CRC} used for predicting and 
visualizing crystal structures.  Crystal ionic radii are based on 
experimental crystal structure determinations, empirical
relationships, and theoretical calculations.  As Fig.~11 shows, they
are in fair agreement with their Slater's counterparts.  We hasten to 
mention that neither Slater's nor the crystal ionic radii are 
expected to coincide with what we define as the size of the ionic core of 
the Rydberg atom, Eq.(\ref{continuity}).  It seems highly plausible, however, that Slater's ionic radii are correlated with the sizes of ionic core of
the Rydberg atom.  

Inspection of Fig.~11 tells us that the average Slater's ionic radius 
slowly grows with $Z^{1/3}$ in fairly good agreement with our
systematic result.  A closer look reveals that our systematic radius 
appears to be consistently smaller than its Slater's counterpart.  If
the same relationship would hold between the systematic and (unknown) 
exact sizes of ionic cores of Rydberg atoms, then the fact that 
experimental and numerical quantum defects in Figs.~9 and 10 are 
generally larger than their systematic counterparts would be explained.       

The large variation of the Slater's ionic radius away from the average trend is
due to the effects of the shell structure.  Their role in determining
the ionic radius can be most easily visualized based on the expression for the 
radius of the maximum electron density (\ref{rmax}) which correlates
with the ionic radius.  This result emphasizes the following main
principles:

(i) As $Z$ increases, all the $n$ levels move down in energy which
amounts to replacing $n$ by its effective counterpart $n^{*} \le n$.  If the
electrons are added to an outer shell, the effective nuclear charge $Z^{*}e$
seen by each of them gradually increases.  This is because the outer
shell electrons are relatively inefficient in shielding the nuclear 
charge.  As a result the ion slowly contracts.       

(ii)  As a new outer shell begins to fill, the effect of going into
the higher shell outweighs the effect of lowering of an $n$ level as
$Z$ increases to $Z + 1$.  This corresponds to an abrupt increase of
the effective principal quantum number $n^{*}$.  Moreover, the
effective charge $Z^{*}e$ seen by the outermost electron drops because
now all remaining $Z - 2$ electrons belong to inner shells thus
efficiently screening the nuclear charge.  These changes in $n^{*}$ and 
$Z^{*}$ cause a sharp increase of the ion size.

Slater's rules \cite{Slater} add a quantitative aspect to these
principles.  In the following explanation of the variation of the ionic radius,
Fig.~11, we are always speaking of the positive singly-charged ions
whose electronic configurations are taken from the NIST database \cite{NIST}.

In going from $^{2}$He ($1s^{1}$), to $^{3}$Li ($1s^{2}$)
the ion size decreases.  As one moves to $^4$Be ([He]$2s^{1}$) the
added electron enters the higher 2$s$ shell, and the ionic radius
sharply increases.  Similar increases
take place as one goes from every alkali to the following alkali earth
ion.  A related jump in ionic radius also occurs past every noble
element ion, $^{29}$Cu ([Ar]$3d^{10}$) $\rightarrow$ $^{30}$Zn 
([Ar]$3d^{10}4s^{1}$),
$^{47}$Ag ([Kr]$4d^{10}$) $\rightarrow$ $^{48}$Cd
([Kr]$4d^{10}5s^{1}$), 
and $^{79}$Au ([Xe]$4f^{14}5d^{10}$) $\rightarrow$
$^{80}$Hg ([Xe]$4f^{14}5d^{10}6s^{1}$) because a higher $s$-shell
starts to be occupied.  An analogous argument explains the sharp increase
of ionic radius while going from $^{24}$Cr ([Ar]$3d^{5}$) to $^{25}$Mn 
([Ar]$3d^{5}4s^{1}$) and from $^{42}$Mo ([Kr]$4d^{5}$) to $^{43}$Tc 
([Kr]$4d^{5}5s^{1}$)

As one moves from $^4$Be ([He]$2s^{1}$) to $^{11}$Na
([He]$2s^{2}2p^{6}$) the $2s$ and $2p$ shells are filled by the
electrons, and the ion size gradually decreases. A similar effect
explains the decrease of the ionic radius along the $^{12}$Mg
([Ne]$3s^{1}$)--$^{19}$K ([Ne]$3s^{2}3p^{6}$), $^{30}$Zn 
([Ar]$3d^{10}4s^{1}$)--$^{37}$Rb ([Ar]$3d^{10}4s^{2}4p^{6}$),
$^{48}$Cd ([Kr]$4d^{10}5s^{1}$)--$^{55}$Cs
([Kr]$4d^{10}5s^{2}5p^{6}$), and $^{80}$Hg ([Xe]$4f^{14}5d^{10}6s^{1}$)--$^{87}$Fr ([Xe]$4f^{14}5d^{10}6s^{2}6p^{6}$) sequences.

The ionic radius decreases through the $^{20}$Ca ([Ar]$4s^{1}$)--
$^{21}$Sc ([Ar]$3d^{1}4s^{1}$)--$^{22}$Ti ([Ar]$3d^{2}4s^{1}$)
segment. This happens because the electrons filling the
$3d$ shell only partially screen the nuclear charge - as a result
the outer $4s$ electron sees a gradual increase of effective $Z^{*}$.
The same argument explains the decrease of ion size while going from
$^{25}$Mn ([Ar]$3d^{5}4s^{1}$) to $^{26}$Fe ([Ar]$3d^{6}4s^{1}$) which
is merely a continuation of the Ca-Ti segment.  The decrease of ionic
radius through the $^{73}$Ta ([Xe]$4f^{14}5d^{3}6s^{1}$)--$^{77}$Ir   
([Xe]$4f^{14}5d^{7}6s^{1}$) series can be similarly understood.  In
fact, the first entry of this series is $^{70}$Yb
([Xe]$4f^{14}5d^{0}6s^{1}$) where we intentionally modified the
standard notation to show the absence of the $5d$ electron.

The size of the ion first abruptly decreases while going from $^{22}$Ti 
([Ar]$3d^{2}4s^{1}$) to $^{23}$V ([Ar]$3d^{4}$) and then continues
decreasing more gradually as one moves to $^{24}$Cr ([Ar]$3d^{5}$).
The sudden change is due to the fact that the outer shell changes from
$4s$ to $3d$ which can be viewed as a decrease in the effective
quantum number $n^{*}$.  The subsequent slower increase of the ionic
radius is due to the increase of the effective nuclear charge $Z^{*}e$
seen by the larger number of $3d$ electrons.  The same trend is
exhibited in the $^{40}$Zr ([Kr]$4d^{2}5s^{1}$)--$^{41}$Nb
([Kr]$4d^{4}$)--$^{42}$Mo ([Kr]$4d^{5}$) sequence.  A very similar
behavior is found in the $^{26}$Fe ([Ar]$3d^{6}4s^{1}$)--$^{27}$Co
([Ar]$3d^{8}$)--$^{28}$Ni ([Ar]$3d^{9}$)--$^{29}$Cu ([Ar]$3d^{10}$), 
$^{43}$Tc ([Kr]$4d^{5}5s^{1}$)--$^{44}$Ru ([Kr]$4d^{7}$)--
$^{45}$Rh ([Kr]$4d^{8}$)--$^{46}$Pd ([Kr]$4d^{9}$)--$^{47}$Ag 
([Kr]$4d^{10}$), and $^{77}$Ir ([Xe]$4f^{14}5d^{7}6s^{1}$)--
$^{78}$Pt ([Xe]$4f^{14}5d^{9}$)--$^{79}$Au ([Xe]$4f^{14}5d^{10}$)
series. 

Superficially, a similar steep decrease of the ion size is
followed by more gradual decrease in the $^{70}$Yb
([Xe]$4f^{14}6s^{1}$)--
$^{71}$Lu ([Xe]$4f^{14}6s^{2}$)-- $^{72}$Hf
([Xe]$4f^{14}5d^{1}6s^{2}$) sequence.  This behavior can be explained
by noticing that in both steps the effective $Z^{*}$ felt by a $6s$
electron increases; the increase during the second step is smaller
because inner shell $d$ electrons are more efficient in screening the
nuclear charge than $s$ electrons.  

While going from $^{38}$Sr ([Kr]$5s^{1}$) to $^{39}$Y ([Kr]$5s^{2}$)
and to $^{40}$Zr ([Kr]$4d^{2}5s^{1}$) a decrease of ionic radius
follows by an increase. This happens because the effective $Z^{*}$
felt by an outer $5s$ electron first increases and then decreases.
The increase of the size of the ion is somewhat smaller than the
decrease because two $4d$ electrons in Zr only partially shield two
extra units of the nuclear charge. 

For the $^{63}$Eu ([Xe]$4f^{7}6s^{1}$)--$^{64}$Gd
([Xe]$4f^{7}5d^{1}6s^{1}$)--
$^{65}$Tb ([Xe]$4f^{9}6s^{1}$) sequence the ionic radius first
decreases and then increases back to its initial value.  This happens
because the $d$ electron in Gd is less effective in shielding the nuclear
charge than the $f$ electron in Tb (considered perfect in Slater's
scheme).

One of the less intuitive increases of ionic radius takes place while going
from $^{56}$Ba ([Xe]$6s^{1}$) to $^{57}$La ([Xe]$5d^{2}$).  On one
hand, the effective principal quantum number $n^{*}$ decreases which 
according to (\ref{rmax}) should lower the ionic radius.  However as
compared to the $6s$ electron of Ba, the effective $Z^{*}$ seen by one of La's
$5d$ electrons also decreases.  This happens because the inner shell 
$5sp$ electrons screen the nuclear charge more effectively if the
outer shell electrons are in a $d$ state (La) as compared to an $s$
state (Ba). As a result the decrease in
$Z^{*}$ outweighs the decrease in $n^{*}$ thus leading to an
increase of the ion size.  A very similar argument explains the reversed
decrease of ionic radius taking place as one goes from $^{58}$Ce
([Xe]$4f^{1}5d^{2}$) (whose size is identical to that of La) to 
$^{59}$Pr ([Xe]$4f^{3}6s^{1}$) (identical in size to Ba).  Here the role of
the $4f$ electrons merely reduces to compensating for the increase of the
nuclear charge.

The ionic radius does not change as one moves from $^{57}$La
([Xe]$5d^{2}$) to $^{58}$Ce ([Xe]$4f^{1}5d^{2}$) because the increase
of nuclear charge is exactly compensated by adding a $4f$ electron.
The same argument explains the constancy of the ionic radius along
most of the lanthanide sequence, $^{59}$Pr ([Xe]$4f^{3}6s^{1}$)--
$^{63}$Eu ([Xe]$4f^{7}6s^{1}$), $^{65}$Tb ([Xe]$4f^{9}6s^{1}$)--
$^{70}$Yb ([Xe]$4f^{14}6s^{1}$).  The only exception from this trend,
$^{64}$Gd ([Xe]$4f^{7}5d^{1}6s^{1}$, has already been discussed.

Now when the variations of the Slater's ionic radius are understood,
we can compare them with experimental and numerical modulations of the
quantum defect.  Since the numerical data are more extensive than experimental
findings, and the agreement between the two is fairly good, in Fig.~11
we only show the variation of the numerically evaluated quantum defect 
$\delta\mu = \mu - \mu_{sc}$.  The inspection of Fig.~11 leaves no
doubt that the variations of the quantum defect with $Z^{1/3}$ are     
correlated with those of the Slater's ionic radius - even the
minute changes of the latter find their way in the corresponding changes of the
former.

There are however two places where it appears there is a disagreement
with expectation:

(i)  Along most of the lanthanide sequence the Slater's ionic radius does
not change while the variation of the quantum defect decreases.  This
can be understood as an artifact of the Slater's rules.  In reality
the $f$ electrons do not perfectly screen the nuclear charge.
Therefore the effective charge seen by the outer $6s$ electron
increases with $Z^{1/3}$ and correspondingly the ionic radius
should decrease.  Then the quantum defect variation should decrease as well
which is in correspondence with numerical results.  We also note that the
Gd dip of the ionic radius is reproduced in the quantum defect variation.

(ii)  For the elements past Hg, the Slater's ionic radius decreases
with $Z^{1/3}$ while the variation of the quantum defect increases.
The reason why it happens is unclear.  It cannot be ruled out that
here the effects of the shell structure might be so strong that our
correlation rule derived from semiclassical arguments breaks down 
qualitatively.

Overall, the analysis of this section makes it certain that the gross
features of the quantum defect variation with $Z^{1/3}$ are due to the 
effects of the shell structure.     

\subsection{Zel'dovich modulation}

Qualitative analysis is of little use in trying to see the Zel'dovich
effect in experimental and numerical data because the Zel'dovich
modulation has an amplitude which is several times smaller than that
due to the effects of the shell structure (see Fig.~10).  A way to
proceed quantitatively is suggested by the fashion in which the shell and
systematic effects are coupled.   

Inspecting the limiting expression for the Zel'dovich modulation,  
Eq.(\ref{zel'oscillation}), it is straightforward to see that after 
replacing systematic $U_{s}(r)$ with the one accounting for the shell 
structure, the amplitude of the Zel'dovich modulation will become 
strongly sensitive to the effects of the shell structure because it is
determined by the logarithmic derivative of the inner potential
$U_{s}(r)$ and the range parameter $x_{0}$.  On the other hand, the
period of the oscillation is far less sensitive to the effects of the
shell structure since it is determined by the number of de Broglie's
half-waves fitting inside the ionic core of the atom.

This last observation suggests that it might be possible to see the
Zel'dovich effect in the experimental and numerical Fourier spectra of
the quantum defect variation $\delta\mu = \mu - \mu_{sc}$ (see
Fig.~10) as a peak whose location can be brought in correspondence
with the systematic theory.  To proceed in this direction, in a range of $Z^{1/3}$ of length $L$ we
expand the quantum defect variation into a Fourier series
\begin{equation}
\label{Fourier}
\delta\mu(Z^{1/3}) = \sum \limits_{k=2\pi p/L} \mu_{k} \exp(ikZ^{1/3})
\end{equation}     
where $p = 0, \pm 1, \pm 2, ...$.  In order to numerically evaluate
the Fourier coefficients $\mu_{k} = \mu^{*}_{-k}$, the experimental and
numerical $\delta\mu(Z^{1/3})$ dependences (see Fig.~10) were fitted
with a cubic spline which was then sampled equidistantly in $Z^{1/3}$
to extract the Fourier spectrum.  The result for the magnitude of the
Fourier coefficients $|\mu_{k}|$ as a function of $k$ is displayed in
Fig.~12 as a series of solid dots 
\begin{figure}
\includegraphics[ width=1.0\columnwidth,
keepaspectratio]{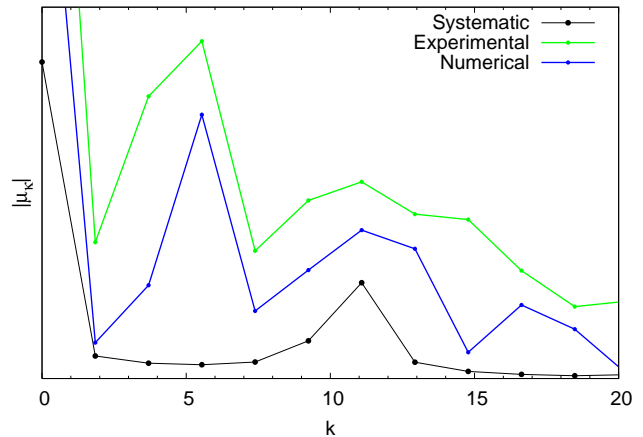}

\caption{Systematic, experimental and numerical amplitudes of the 
Fourier coefficients of the quantum defect
variation (arbitrary units, same normalization ) $|\mu_{k}|$ as
functions of $k$ for $k \ge 0$.  The peaks at $k \simeq 11$ correspond
to the Zel'dovich effect.}

\end{figure}
which for convenience are connected by straight line segments. The
uncertainty of the location of each dot along the $k$ axis,
$2\pi/L$, is the distance between the nearest values of $k$.  Since
the last available experimental quantum defect corresponds to
$^{83}$Bi (see Table 1), for both experimental and numerical data we
restricted ourselves to $L = 83^{1/3}$.  The finite values of $\mu_{0}$
correspond to the presence of nonzero background in experimental and
numerical $\delta\mu(Z^{1/3})$ dependences and are of no interest to us.   

For comparison in Fig.~12 we also show the Fourier spectrum of our systematic
calculation which, as expected, has only one peak corresponding to the
Zel'dovich effect.  The position of the peak along the $k$ axis can be
understood from the large $Z$ asymptotics of the Thomas-Fermi action 
$S_{0}/\hbar \simeq 5.2Z^{1/3}$ (\ref{TFaction}).  Comparing this with
the limiting expression for the Zel'dovich modulation
(\ref{zel'oscillation}) we would expect a peak at $k \simeq 10.4$.
The peak in Fig.~12 is located at a slightly different value of $k \simeq 11$
which is due to the fact that the asymptotic behavior $S_{0}/\hbar \simeq
5.2Z^{1/3}$ becomes numerically accurate only for $Z^{1/3}$ exceeding
$10$.
    
Both the experimental and numerical spectra in Fig.~12 have peaks at
the same value of $k \simeq 11$ which we argue are the signatures of
the Zel'dovich effect. 

It is curious that both experimental and numerical spectra have another
peak in common located at $k \simeq 5.5$.  This peak which is about twice as
high as that due to the Zel'dovich effect is natural to relate to
the effects of the shell structure.  The existence of this peak
translates into the $(2\pi/5.5)Z^{1/3} = 1.14Z^{1/3}$ periodicity of
the quantum defect variation due to the effects of the shell
structure.  This conclusion resembles the $Z^{1/3}$ periodic
oscillation of the ground-state energy of an atom away from
the systematic trend \cite{Englert}.  With uncertainty of the peak location
in mind, one may speculate that our result is a manifestation of the 
same effect for highly-excited states.  More work is necessary to
bring understanding to this issue.    

We also repeated the same Fourier analysis by choosing the range of
$Z^{1/3}$ to be $L = 102^{1/3}$ which includes all the numerical
data.  As far as experimental data go in the range of $Z^{1/3}$
between $83^{1/3}$ and $102^{1/3}$ we used extrapolation of our cubic
spline fit.  As a result the peaks just discussed slightly change
their positions and amplitudes but within the $2\pi/L$ uncertainty
systematic, experimental and numerical Fourier spectra share a peak
in common corresponding to the Zel'dovich effect.  Similarly,
experimental and numerical spectra continue to share a peak in common
due to the effects of the shell structure. 

\section{Conclusions and future directions}

In this paper we analyzed in a model-independent fashion the weakly-bound
$s$ spectra of the distorted Coulomb problem for arbitrary
relationship between the range of the inner potential and 
Bohr's radius of the Coulomb field.  We demonstrated that the spectra
are fairly sensitive to the binding properties of the inner potential
which constitutes the essence of the Zel'dovich effect, and established 
the corresponding details of spectral changes.  Armed with these
results, we conducted an analysis of experimental and numerical Rydberg
spectra of atoms along the Periodic Table which indeed show an evidence of
the Zel'dovich effect.  Our analysis can be extended and adopted in several
directions:

First, there is an abundance of experimental and numerical data for atomic
Rydberg states of finite angular momenta  which are likely to contain
signatures of the Zel'dovich effect.  However, in the limit of a very 
short-ranged inner potential the way the effect manifests itself is
somewhat different from its $s$ state counterpart \cite{Popov2}.  This
observation makes it pertinent to generalize our analysis to the case
of finite angular momentum.

It has been known for some time \cite{Resca} that the Rydberg formula
(\ref{Rydberg}) is superior to the Wannier (Bohr) formula quoted in 
textbooks \cite{AM} in representing excitonic spectra in condensed matter 
systems.  Experimental examples here include clean and doped rare-gas
solids and rare-gas impurities in solid hydrogen.  Although this is a
context in which the Zel'dovich effect has been originally discovered
\cite{Zel'dovich}, to the best of our knowledge there were no attempts
to relate it to excitonic quantum defects.  Because of the dielectric
screening of the Coulomb interaction, the Zel'dovich effect in these
systems is expected to be more pronounced than in atomic Rydberg
spectra.  Only minor changes to our analysis are needed to understand
the excitonic Rydberg spectra.

Other examples of systems where the Zel'dovich effect should have
experimental signatures include Rydberg ions and electronic image states
\cite{image}.   

\section{ACKNOWLEDGMENTS}

We are grateful to T. F. Gallagher and R. R. Jones for sharing with
us their expertise. We also acknowledge R. M. Kalas and X. Yang who
participated in the initial stages of this project. This work was
supported by the Chemical Sciences, Geosciences and Biosciences Division,
Office of Basic Energy Sciences, Office of Science, U. S. Department
of Energy.


\begin{thebibliography}{10}
\bibitem{note1}A more accurate calculation based on Eq.(\ref{qdgeneral}) gives $\mu=-2a_{s}/a_{B}-x_{0}^{4}/32$
which is negative for sufficiently weak $U_{s}(r)$; for $x_{0}\ll1$
this effect however is negligible as can be seen from Fig.~3.
\bibitem{note2}Eq.(\ref{zel'oscillation}) predicts that there is an unphysical divergence
as $Z\rightarrow1+0$ which is present but hardly noticeable in the
inset. The oscillation shown in bold is everywhere finite.
\end{thebibliography}
\end{document}